\documentclass[twocolumn,trackchanges]{aastex631}
\usepackage{amssymb}
\usepackage{graphicx}
\usepackage{grffile}
\usepackage{epsfig}
\usepackage{epstopdf}
\usepackage[T1]{fontenc}
\usepackage{url}
\usepackage{savesym}
\savesymbol{tablenum}
\usepackage{siunitx}
\restoresymbol{SIX}{tablenum}
\usepackage{newtxtext,newtxmath}
\usepackage[T1]{fontenc}

\DeclareRobustCommand{\VAN}[3]{#2}
\let\VANthebibliography\thebibliography
\def\thebibliography{\DeclareRobustCommand{\VAN}[3]{##3}\VANthebibliography}

\usepackage{graphicx}	% Including figure files
\usepackage{amsmath}	% Advanced maths commands
\usepackage{amssymb}	% Extra maths symbols
\usepackage{amsmath}
\usepackage{siunitx}
\usepackage{tablefootnote}

\usepackage{url}
\usepackage{hyperref}
\usepackage{cleveref}
\usepackage[flushleft]{threeparttable}

\def\ga{\mathrel{\raise.3ex\hbox{$>$\kern-.75em\lower1ex\hbox{$\sim$}}}}
\def\la{\mathrel{\raise.3ex\hbox{$<$\kern-.75em\lower1ex\hbox{$\sim$}}}}

\shorttitle{UV spectrum of J1059+4251}
\shortauthors{A.\ Citro et al.}

\turnoffeditone

\hypersetup{linkcolor=purple,citecolor=teal,filecolor=cyan,urlcolor=purple}

\begin{document}

\title{\large \textbf{SDSS\,J1059+4251, a highly magnified \boldmath$z \sim2.8$\unboldmath\ star-forming galaxy:\\ ESI observations of the rest-frame UV spectrum}}

\correspondingauthor{Annalisa Citro}
\email{acitro@uwm.edu}

\author[0000-0003-3948-6688]{Annalisa Citro}
\affil{The Leonard E.\ Parker Center for Gravitation, Cosmology and Astrophysics, Department of Physics,\\ University of Wisconsin-Milwaukee, 3135 N Maryland Avenue, Milwaukee, WI 53211, USA}

\author[0000-0001-9714-2758]{Dawn K. Erb}
\affiliation{The Leonard E.\ Parker Center for Gravitation, Cosmology and Astrophysics, Department of Physics,\\ University of Wisconsin-Milwaukee, 3135 N Maryland Avenue, Milwaukee, WI 53211, USA}

\author[0000-0002-5139-4359]{Max Pettini}
\affiliation{Institute of Astronomy, University of Cambridge, Madingley Road, Cambridge, CB3 0HA, UK}

\author{Matthew W. Auger}
\affiliation{Institute of Astronomy, University of Cambridge, Madingley Road, Cambridge, CB3 0HA, UK}

\author[0000-0003-2344-263X]{George D. Becker}
\affiliation{Department of Physics \& Astronomy, University of California, Riverside, CA 92521, USA}

\author[0000-0003-4372-2006]{Bethan L. James}
\affiliation{AURA for ESA, Space Telescope Science Institute, 3700 San Martin Drive, Baltimore, MD 21218}

\begin{abstract}
Detailed analyses of high-redshift galaxies are challenging due to their faintness, but this difficulty can be overcome with gravitational lensing, in which the magnification of the flux enables high signal-to-noise ratio (S/N) spectroscopy. 
We present the rest-frame ultraviolet (UV) Keck Echellette Spectrograph and Imager (ESI) spectrum of the newly discovered  $z=2.79$ lensed galaxy SDSS J1059+4251. With an observed magnitude $\rm{F814W} = 18.8$ and a magnification factor $\mu= 31\pm 3$, J1059+4251 {is both highly magnified and intrinsically luminous, about two magnitudes brighter than $M_{\rm UV}^*$ at $z\sim2$--3.
With stellar mass $M_*= (3.22\pm 0.20) \times10^{10}\, \rm M_{\odot}$, star formation rate $\rm SFR=50\pm 7\, \rm M_{\odot}\,yr^{-1}$, and stellar metallicity $Z_* \simeq 0.15-0.5\, Z_{\odot}$}, J1059+4251 is typical of bright star-forming galaxies at similar redshifts. 
Thanks to the high S/N and the spectral resolution of the ESI spectrum, we are able to separate the interstellar and stellar features and derive properties that would be inaccessible without the aid of the lensing. We find evidence of a gas outflow with speeds up to $\rm - 1000\,km\,s^{-1}$, and of an inflow that is probably due to accreting material seen along a favorable line of sight. 
We measure relative elemental abundances from the interstellar absorption lines and find that $\alpha$-capture elements are overabundant compared to iron-peak elements, suggestive of rapid star formation. 
However, this trend may also be affected by dust depletion. Thanks to the high data quality, our results represent a reliable step forward in the characterization of typical galaxies at early cosmic epochs.

\end{abstract}



%%%%%%%%%%%%%%%%%%%%%%%%%%%%%%%%%%%%%%%%%%%%%%%%%%

%%%%%%%%%%%%%%%%% BODY OF PAPER %%%%%%%%%%%%%%%%%%

\section{Introduction}

\begin{figure*}
	\includegraphics[width=\columnwidth]{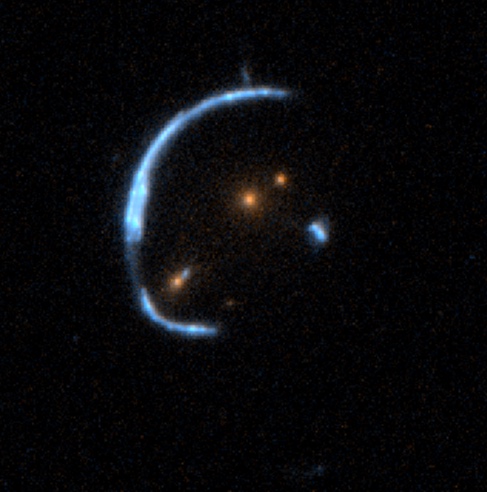}~~~~~\includegraphics[width=\columnwidth]{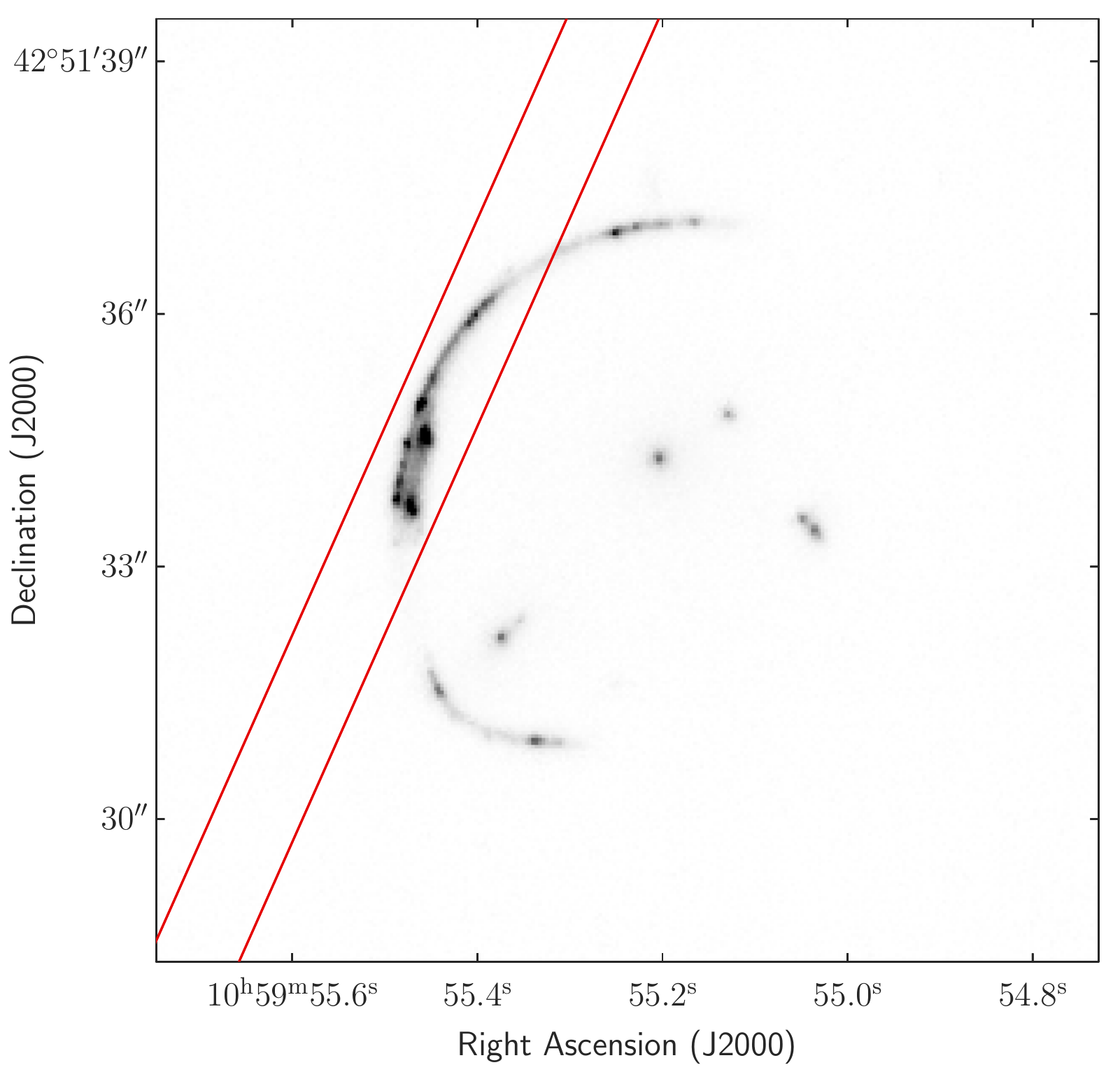}
      \caption{Left: Color composite image of SDSS\,J1059+4251, obtained by combining \textit{Hubble Space Telescope} (\textit{HST}) Wide Field Camera 3 (WFC3) images in the F606W and F814W filters. The top of the figure corresponds to the north direction, while left corresponds to east. Right: F814W image showing the positioning of the ESI slit used for the observations reported in this paper.}
   \label{fig:J1059_image}
\end{figure*}

In the last two decades or so, the number of galaxies discovered at redshifts $z \simeq 1$--10 has increased spectacularly thanks largely to the development of appropriate colour selection criteria, pioneered by \cite{Steidel+1996} (see also \citealt{Adelberger+2004}), and reviewed by \citet{Shapley2011} (see also \citealt{Madau_Dickinson2014}).
However, galaxies at high redshifts are faint ($m_{\rm R}^\ast = 24.4$ at $z = 2-3$; \citealt{Steidel+1999}; \citealt{Reddy+2008}) and the signal-to-noise ratio (S/N) needed to enable detailed spectroscopic studies is hard to achieve, until the next generation of 30+ m optical/IR telescopes comes on line. Besides long integration times (e.g.\ \citealt{Erb+2010}) and stacking techniques (e.g.\ \citealt{Zhu+2015}, \citealt{Steidel+2016}, \citealt{Rigby+2018a}, \citeyear{Rigby+2018b}), another effective way to overcome these difficulties is to study high redshift gravitationally lensed galaxies, where the magnification furnished by the gravitational lensing provides high S/N spectra which can be analyzed in greater detail. In particular, rest-frame ultraviolet wavelengths are especially interesting, since they encode meaningful information about the star formation processes that took place in the early Universe. UV stellar continua can be used to infer the galaxy luminosity, star-formation rate and dust extinction (e.g.\ \citealt{Kennicutt1998}, \citealt{Calzetti+2000}, \citealt{Salim+2007}, \citealt{Wilkins+2012b}), while
individual spectral features (in absorption and in emission) provide valuable information on both the galaxy stellar populations and
the interstellar medium (ISM) (e.g. \citealt{Shapley+2003}, \citealt{Berry+2012}, \citealt{Talia+2012}, \citealt{Steidel+2016}, \citealt{Du+2018}, \citealt{Cullen+2019}), including the chemical composition of young OB stars and the ISM from which these stars have recently formed. Furthermore, the relative
abundances of elements produced on different timescales (i.e. iron-peak vs.\ 
$\alpha$-capture elements) give clues about the galaxy's past history of star formation, as shown by the recent optical study of SDSS galaxies by \citet{Gallazzi+2021}. However, the ability to recover accurately all of these properties relies on the acquisition of high quality spectra of sufficiently high S/N and resolution to recognize stellar
spectral lines against the continuum and reliably separate them from interstellar features.

An increasing number of gravitationally lensed galaxies have been spectroscopically analyzed in the UV range in the past years. One of the best-known sources is MS 1512-cB58 \citep{Pettini+2000,Pettini+2002}.
 Some other well-known cases are the Lynx arc \citep{Fosbury+2003}, BD38 \citep{Dow-Hygelund+2005}, the "Cosmic Horseshoe" \citep{Quider+2009, James+2018}, the "8 o'clock arc" \citep{Dessauges-Zavadsky+2010}, the "Cosmic Eye" \citep{Quider+2010}, and more recently SGAS J105039.6+001730 \citep{Bayliss+2014}, Cassowary20 \citep{Pettini+2010, James+2014}, J1110+6459 \citep{Rigby+2017} and SL2S J021737–051329  \citep{Berg+2018}. Through the study of gravitationally lensed sources it appears that the population of high redshift galaxies is characterized by low metallicities ranging from 1/20 \citep{Berg+2018} to 1/2 $Z_{\odot}$ (\citealt{Quider+2009}, \citeyear{Quider+2010}), ages of a few hundreds of Myrs (\citealt{Pettini+2002}, \citealt{Dessauges-Zavadsky+2010}) and high ionization levels $ -3 \lesssim \log (U) \lesssim -1.5$ (\citealt{Hainline+2009}, \citealt{Richard+2011}, \citealt{Berg+2018}). Moreover, the observed overabundance of metals produced on shorter timescales by Type II supernovae (SNII)  with respect to those produced on longer timescales by Type Ia supernovae (SNIa) suggests the presence of rapid star formation timescales. Evidence of large-scale outflows driven by stellar winds and supernovae have been found in most gravitationally lensed galaxies (e.g.\ \citealt{Pettini+2002}, \citealt{Bayliss+2014}), and in rare cases also indications of possible inflows of material have been observed. The latter might be produced by gas which has been previously ejected and is now falling back on to the galaxy (e.g.\ \citealt{Quider+2010}).
 The majority of the studies performed on gravitationally lensed galaxies have involved sources which are magnified by factors $< 25$, while only a few objects with magnification factors $> 30$ have been spectroscopically investigated at {ultraviolet and optical wavelengths (\citealp{Pettini+2002}, \citealp{Ebeling+2009} , 
 \citealp{VanDerWel+2013}, \citealt{Bayliss+2014}, \citealt{Rigby+2017}, \citealt{Rivera-Thorsen+2017}).} Yet, the most highly magnified sources are certainly the most interesting ones, since they potentially provide the highest possible S/N and thus the most accurate results.

\begin{figure*}
	\includegraphics[width=\textwidth]{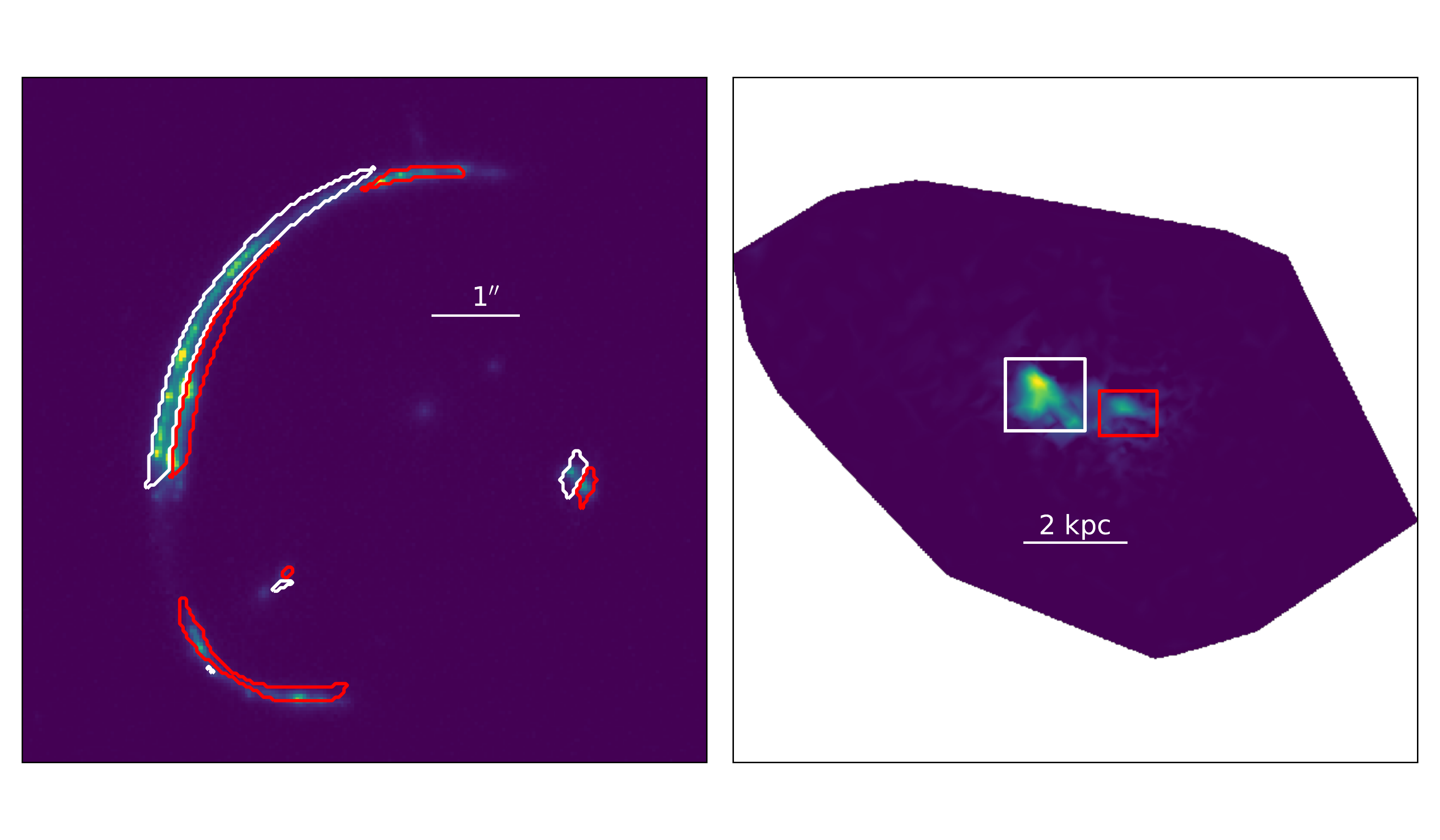}
	\caption{Source plane reconstruction with our lensing model, described in Section~\ref{sec:LensModel}. Left: J1059+4251 in the image plane seen through the WFPC3 F606W filter; the colour contours correspond to portions of the arc that map onto different regions in the source. Right: J1059+4251 in the source plane using the same colour key as in the left panel. {The characteristic residuals when subtracting the model from the observed arc are $\sim0.1$ dex in the optical bands and $\sim0.03$ dex in the NIR bands (due to the lower resolution).}} 
   \label{fig:J1059_LensModel}
\end{figure*}

In this paper, we present the first results from a concerted study of a newly discovered, highly magnified, high redshift gravitationally lensed galaxy, J1059+4521 (J1059, hereafter). 
This galaxy has been identified in the course of a search for lensed QSOs, using SDSS \textit{ugriz} and WISE photometry \citep{Wright+2010}. J1059 has a redshift $z=2.8$ (see Section \ref{sec:J1059zStars}) and is magnified by a galaxy complex at $z\sim 0.7$ by a factor $\mu= 31\pm 3$, which is one of the highest magnifications known to date (similar to that of MS 1512-cB58; \citealp{Seitz+1998}).

The aim of this paper is to define the average physical and evolutionary properties of J1059 by means of its stellar and interstellar UV spectral features, and to add another piece of knowledge to the characterization of the galaxy population at high redshift. It is worth noting that gravitationally lensed galaxies are also opening the way to spatially resolved studies. Their lensing magnification provides high resolution on sub-kpc-scales, enabling us to spatially resolve the internal kinematics (e.g. \citealt{Swinbank+2009}, \citealt{Jones+2010},  \citealt{Stark+2013}, \citealp{James+2018}). We will focus on this kind of analysis in a companion paper (James et al. 2021 in prep.).

The current paper is organized as follows. In Section \ref{sec:LensModel} we describe the lensing model and the morphology of J1059; in Section \ref{sec:obs} we explain how the J1059 photometry and the UV spectroscopy were obtained, while in Section \ref{sec:J1059zStars} we derive the systemic redshift. In Section \ref{sec:general_prop} we infer stellar population properties of J1059 such as the stellar mass and the star formation rate; in Section \ref{sec:stellar_pop} and \ref{sec:ISM} we analyze the interstellar spectrum of J1059, including the gas kinematics and the column densities of neutral hydrogen and nine metal ions, while in Section \ref{sec:chemical_comp} we discuss the chemical composition of the interstellar gas. Lastly, in Section \ref{sec:conc} we summarize our findings.

 We adopt throughout a $\Lambda$CDM cosmology with $H_0 = 70$\,km~s$^{-1}$~Mpc$^{-1}$, $\Omega_{\rm m}= 0.3$, and $\Omega_{\Lambda}= 0.7$. All distances are given in physical (proper) units unless stated otherwise. We use the abbreviations ppc and pkpc to indicate physical units of parsecs and kiloparsecs. At the redshift of J1059
($z = 2.8$), 1 arcsec on the sky corresponds to 7.86\,pkpc in the image plane. Moreover, throughout the paper, we consider high redshifts the range $z=2-3$.

\section{SDSS J1059+4251: one of the brightest galaxy-scale lensed galaxies identified to date}
\label{sec:LensModel}

SDSS\,J1059+4251  was identified in the course of a search for lensed QSOs utilizing Sloan Digital Sky Survey (SDSS) and Wide-field Infrared Survey Explorer (WISE) photometry, due to its WISE $\textrm{W1}-\textrm{W2}\sim0.5$ color and multiple detections in SDSS \citep{Lemon2018}.
The system consists of a blue background galaxy at redshift $z = 2.8$, gravitationally
lensed into a bright arc and fainter counterimages by a foreground sparse group of galaxies
whose colors suggest that they are at $z \sim 0.7$. The arc is unusually bright, 
with $g$, $r$, and $i$ magnitudes of 19.3, 18.7, and 18.5 respectively. These are $\sim 5$ times
brighter than MS1512-cB58 (AB$_{6540} = 20.41$ at $z = 2.72$, \citealt{Ellingson1996}),
which remains one of the best studied high redshift galaxies.

\textit{HST} images are reproduced in Figure \ref{fig:J1059_image}. As can be appreciated from the figure, there is considerable structure in the lensed arc, with several bright knots indicating that the background galaxy has a complex morphology. To reconstruct the source image, we followed the adaptive pixellated source modeling technique described by \cite{VegettiKoopmans2009}. {Our lens model is based on the F606W data, but we verified that adding the other bands does not improve the fit and that the F606W-based model reproduces very well the data in the other bands.}

The intrinsic source surface brightness distribution was described on an irregular grid of pixels that approximately follows the magnification of the lens, with a PSF-deconvolved intensity at each pixel determined from the lens mass model and observed data. We model the lensing mass distribution as two singular isothermal models, one centered on the central galaxy and the other on the galaxy to its southeast (see the left panel of Figure \ref{fig:J1059_image}); the former is modelled as an ellipsoid and the latter as a sphere, and we also include an external shear. {The initial models also included the galaxy visible at about 1\arcsec\ to the northwest of the main lens. However, since its resulting amplitude was consistent with zero, we excluded it  from the final modeling to reduce the size of the parameter space.
}

The reconstructed source surface brightness distribution can be seen in the right-hand panel of Figure \ref{fig:J1059_LensModel}. It shows an elongated structure with two main concentrations of stellar light, separated by $\sim 1.5$ pkpc, contributing to the lensed image.
This morphology is not unusual for star-forming galaxies at $z \sim 2$--3, although without the aid of gravitational lensing it is normally seen on larger scales (\citealp{Law+2007}, \citealp{Conselice2014}, but see also  \citealp{Johnson+2017}).

Our lensing model returns a remarkably high magnification for this system, $\mu= 31\pm3$ which is, to our knowledge, one of the highest magnifications provided by a galaxy-scale lens. {We also note that due to differential magnification across the extended source the \textit{spectroscopic features} may experience different magnification than the flux observed in the broadband imaging. This cannot be quantified with the data analyzed in this paper, but will be investigated with spatially resolved spectroscopy in the future (James et al, in preparation).}

\section{Observations and data reduction}
\label{sec:obs}

\begin{figure*}
\centering
  \includegraphics[width=1\linewidth]{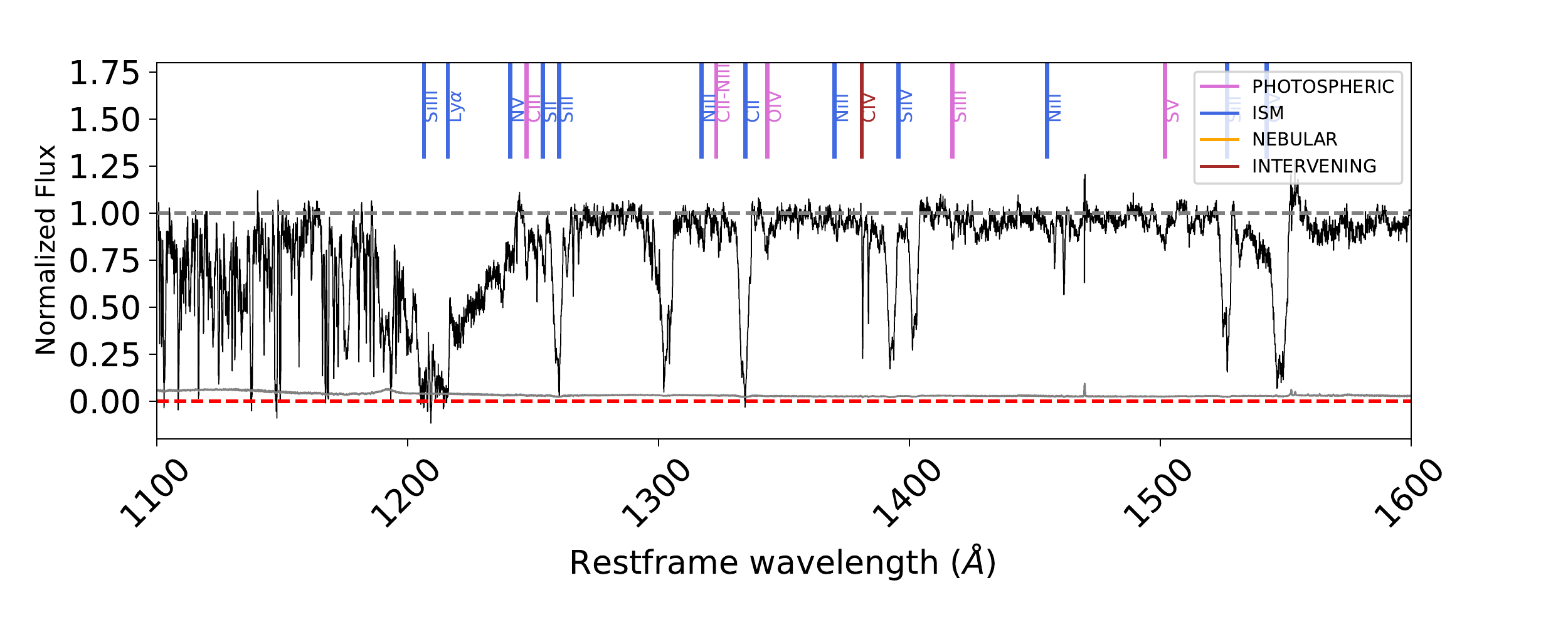}\\
  \includegraphics[width=1\linewidth]{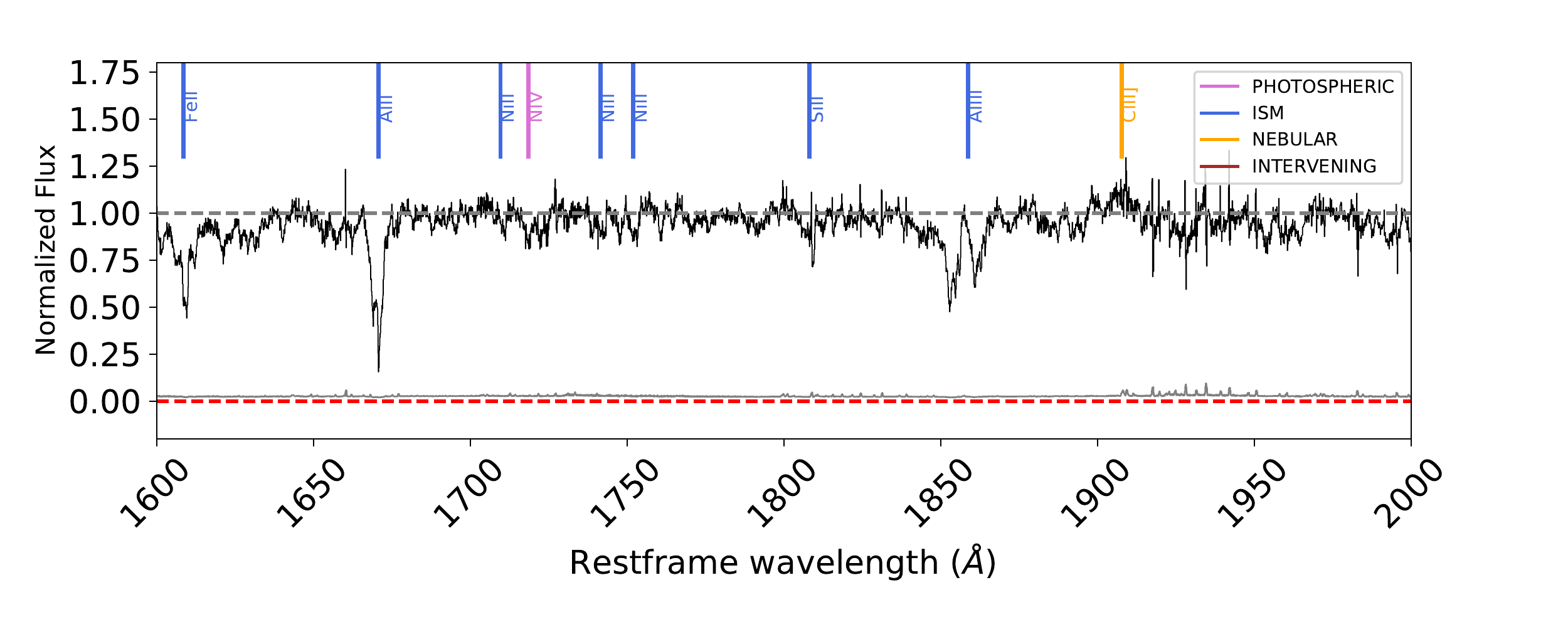}
\caption{Normalized ESI spectrum of J1059 in the wavelength range 1100—2000\,\AA. Vertical lines identify some of the most important spectral features, colour-coded according to their origin, as indicated in the panel. The error spectrum is shown in grey.}
\label{fig:ESI_spec}
\end{figure*}

Prompted by the unusual brightness of the lensed arc in J1059, we have targeted this system with a variety of observations, including \textit{HST} imaging, slit spectroscopy with the Echellette Spectrograph and Imager \citep[ESI;][]{Sheinis+2000} on the Keck\,II telescope, and integral field spectroscopy with the Keck Cosmic Web Imager \citep[KCWI;][]{Morrissey+2018}, also on the Keck\,II telescope. The KCWI observations are the subject of a forthcoming paper (James et al, in preparation), although we use some of their findings here. This paper focuses primarily on the ESI spectroscopy, after a brief description of the \textit{HST} imaging. 

\subsection{\textit{HST} and \textit{WISE} Imaging}
\label{sec:HST_im}

We obtained  photometry through the F606W, F814W, F125W and F160W filters of the  Wide Field Camera 3 (WFC3) UVIS/IR imager on the \textit{HST} in order to constrain the star formation and stellar population properties across the source and on sub-kpc scales. 
The F606W observations were acquired on 14 February 2018 (proposal ID 15223, PI Auger), and the F814W, F125W and F160W observations were obtained on 19 February 2019 (proposal ID 15467, PI Erb). The spatial sampling was optimized by using a 4-point box dither for the UVIS observations. For the IR observations with larger pixels, we used a 4-point box dither followed by a 3-point line, for a total of 7 exposures at 6 different positions. The repeated position, corresponding to the first exposure in each dither pattern, enables a consistency check of the slightly different exposure times of the two patterns, and the combination of the two patterns improves the sub-pixel sampling over either pattern alone. We used the STEP100 sampling sequence in order to obtain good sampling on both the lensed arc and on PSF stars to be used for the de-lensing model. The dither sequences also enable rejection of detector artifacts and cosmic rays. The final images in each filter are drizzled onto 0\farcs04 pixels centered on the lensing galaxy using a Lanczos3 kernel for the UVIS images and a Gaussian kernel for the IR images. After finding the lens model using the F606W\footnote{Adding the 
other \textit{HST} bands does not improve the lens model. We note that the model fitted to the F606W data does a very good job at modelling the data in the other bands with well-focused sources.} data (Section \ref{sec:LensModel}), we then model the three foreground lensing galaxies as Sersic profiles and simultaneously solve for the best source surface brightness distribution to effectively model away the lens galaxy light and determine the image-plane source flux (the source and lens flux are well-separated in the UVIS data but there is significant overlap in the IR data). 

WISE W1 and W2 imaging data were also used to extend the spectral energy distribution (SED) to redder wavelengths. Because of the much poorer spatial resolution compared to \textit{HST} (the resolution is approximately 6\arcsec\ in W1 and W2) the light from the background source is significantly contaminated by the light from the lensing galaxies. Nevertheless, if we use the fits to the \textit{HST} imaging data described above as models for the surface brightness distributions of the background source and foreground lensing galaxies, we can attempt to remove the contribution of the foreground galaxy light. In particular, we convolve the model of the lensed background emission with models of the WISE W1 and W2 PSFs, and we do the same for the foreground lensing galaxies. We also do the same to the observed \textit{HST} data of two galaxies immediately to the east and west of the lens system, as their light slightly overlaps with the lensed background emission due to the size of the WISE PSFs. We then find the best amplitudes for each of our four WISE model components (the foreground lensing galaxies, the background lensed emission, and the two galaxies to the east and west) that best fit the observed W1 and W2 images. We find that the foreground lensing galaxies have {very blue W1-W2} colors while the background lensed arc is quite red, consistent with the initial selection of the system as a potential quasar. The \textit{HST} and WISE magnitudes for the total observed light (i.e., in the image plane) of the background galaxy so obtained are listed in Table \ref{tab:table_hst}, together with the total exposure times for each band.

\begin{table}{}
	\centering
\caption{\textit{HST} and WISE photometry on the AB system. {These are the observed magnitudes for the entire lensed source, uncorrected for lensing magnification and uncorrected for foreground Galactic extinction.}}
	\label{tab:table_hst}
		\begin{tabular}{lccc}
		\hline
		   Filter & Magnitude $\pm$ error & Central & Exposure \\
		    & & wavelength (\AA) & time (s)\\
		\hline
	   F606W & 19.07  $\pm$ 0.02 & 5885 & 2572\\
	   F814W & 18.84  $\pm$ 0.02 & 8048 & 2572\\
	   F125W & 18.68  $\pm$ 0.05 & 12486 & 2494\\
	   F160W &  18.37  $\pm$ 0.05 & 15369 &  2494\\
	   WISE1 & 17.59 $\pm$ 0.03 & 33526 & 1480\\
	   WISE2 & 17.50 $\pm$ 0.02 & 46028 & 1440\\
	   \hline
	\end{tabular}
\end{table}

\subsection{ESI Slit Spectroscopy}
\label{sec:ESI_spec}
The ESI observations were conducted on the night between February 27 and 28 2019 UT. We used the 1\,arcsec wide slit at position angle $\rm PA = -24^\circ$ (see Figure~\ref{fig:J1059_image}). The total exposure time was 7 hours, split into 14 1800\,s long integrations. The mean (median) airmass during the observations was 1.27 (1.20) and the seeing was stable at $\sim 0.8$\,arcsec FWHM through the night. 

The data were reduced with a custom pipeline developed by \cite{Becker+2009} and described in some detail in \cite{Lopez+2016}. For each 1800 s exposure, optimal sky subtraction was performed on the unrectified 2D frame following the prescriptions by \cite{Kelson2003} to maximize the S/N ratio and minimize residuals from bright emission lines from the earth's atmosphere. Wavelength calibration (to vacuum heliocentric values) was by reference to comparison spectra produced by Cu-Ar and Hg-Ne-Xe hollow-cathode lamps. Correction for telluric absorption used the atmospheric transmission spectrum of \cite{Hinkle+2003}.

The wavelength and flux calibrated, co-added 1D spectrum covers the wavelength range 3100--10300\,\AA, although the S/N deteriorates rapidly shortwards of $\sim 4075$\,\AA\ and longwards of $\sim 7800$\,\AA; therefore, we limit ourselves to the analysis of this wavelength region, which corresponds to the range 1075--2050\,\AA\ in the rest frame of J1059. The spectral resolution, determined from the widths of sky emission lines, is FWHM\,$\simeq 60$\, km~s$^{-1}$, sampled with three wavelength bins. The S/N in the continuum is between 30 and 40 per 20\,km~s$^{-1}$ ($\sim 0.4$\,\AA) bin over most of the above wavelength range, falling to $15-25$ below 1200\,\AA. 

After transforming the spectrum to the rest-frame of J1059 at $z_{\rm stars} = 2.79556$ (see Section \ref{sec:J1059zStars}), we normalized it by dividing by our best estimate of the stellar continuum following the prescription of \cite{Rix+2004}; this final spectrum is reproduced in Figure \ref{fig:ESI_spec}. It is important to bear in mind throughout the subsequent analysis that the ESI slit (see Figure \ref{fig:J1059_image}) captures the light from \emph{both} regions in the reconstructed source image shown in Figure \ref{fig:J1059_LensModel}. Consequently, all of our findings should be interpreted as approximately average values for the galaxy, while the integral-field spectroscopy of James et al. (in preparation) will assess if and how the galaxy's properties vary with location within the galaxy
(on an unprecedented fine physical scale of $\sim 10$\,ppc).

\section{Systemic Redshift}
\label{sec:J1059zStars}

%\begin{figure}
%	\includegraphics[width=0.49\textwidth]{plot_cfr_sfr_mass_4.pdf}
%      \caption{Comparison between J1059's stellar mass and SFR (obtained from the SED fitting) and the main sequences of star forming galaxies at $z\sim2-4$ derived by \citet{Santini+2017}. The stellar mass and SFR provided by \citet{Santini+2017} have been rescaled from a Salpeter to a Chabrier IMF in order to be consistent with our results.}
%   \label{fig:sfr_mass}
%\end{figure}

As can be appreciated from Figure~\ref{fig:ESI_spec}, the ESI spectrum of J1059 is rich in UV spectral features. The most obvious features are strong interstellar absorption lines (indicated in blue in the two panels of Figure \ref{fig:ESI_spec}) formed in the gas of J1059 located in front of (most of) the early-type stars producing the UV continuum. We interpret the very strong feature centred near 1210\,\AA\ as the blend of a damped Ly$\alpha$ absorption line with mostly redshifted (relative to the stars) Ly$\alpha$ emission (see Section \ref{sec:Lya}). The interstellar lines are seen against a background of low contrast photospheric absorption lines due to OB stars; the earliest spectral types within the stellar population give rise to typical P-Cygni profiles in the high ionization resonance doublets of C\,{\sc iv}~$\lambda\lambda 1548, 1550$, Si\,{\sc iv}~$\lambda\lambda 1393, 1402$ and N\,{\sc v}~$\lambda\lambda 1238, 1242$. Emission lines are weak: apart from weak nebular C\,{\sc iii}]~$\lambda \lambda 1907, 1909$, we do not detect O\,{\sc iii}]~$\lambda\lambda 1661, 1666$, nor any emission lines due to transitions to fine structure levels of the ground states of C\,{\sc ii} and Si\,{\sc ii} \citep{ScarlataPanagia2015} which can sometimes be clearly visible in such spectra \citep[see, for example,][]{Erb+2010}. He\,{\sc ii}~$\lambda 1640$ is also not detected. Evidently, the spectrum of this starburst galaxy is dominated by strong absorption lines. Finally, we detect narrow absorption lines from intervening gas not associated with J1059; apart from the Ly$\alpha$ forest due to the intergalactic medium, a pair of narrow absorption lines near 1380\,\AA\ is identified as a C\,{\sc iv}~$\lambda\lambda 1548, 1550$ doublet at $z_{\rm abs} = 2.38651$.

\begin{table}{}
\centering
	\caption{Photospheric absorption features used to measure the systemic redshift of J1059+4251.}
	\label{tab:table_sysz}
	\begin{center}
	 \addtolength{\leftskip} {-0.8cm}
		\begin{tabular}{lccc}% four columns, alignment for each
		\hline
		Ion & $\lambda_{\rm lab}$ (\AA) & $\lambda_{\rm obs}$ (\AA) & $z_{\rm stars}$~~\,\\
		
		\hline
		C\,{\sc iii} &1247.38 & $4735.25 \pm 0.19$ & $2.79616 \pm 0.00015$ \\
		C\,{\sc ii} &1323.93 & $5026.01 \pm 0.17$ & $2.79628 \pm 0.00013$ \\
	    N\,{\sc iii} &1324.31 & $5026.00 \pm 0.18$ & $2.79517 \pm 0.00013$ \\
	    O\,{\sc iv}  &1343.35 & $5098.24 \pm 0.32$ & $2.79516 \pm 0.00024$ \\
		Si\,{\sc iii} & 1417.24 & $5378.92 \pm 0.13$ &$ 2.79535 \pm 0.00009$ \\
		S\,{\sc v}   & 1501.76 & $5698.20 \pm 0.39$ & $2.79435 \pm 0.00026$  \\
		N\,{\sc iv}  & 1718.55 & $6523.12 \pm 0.31$ & $2.79571 \pm $ 0.00018 \\
		\hline
	\end{tabular}
	\end{center}
\end{table}

\begin{figure*}
\centering
   \includegraphics[width=1\linewidth]{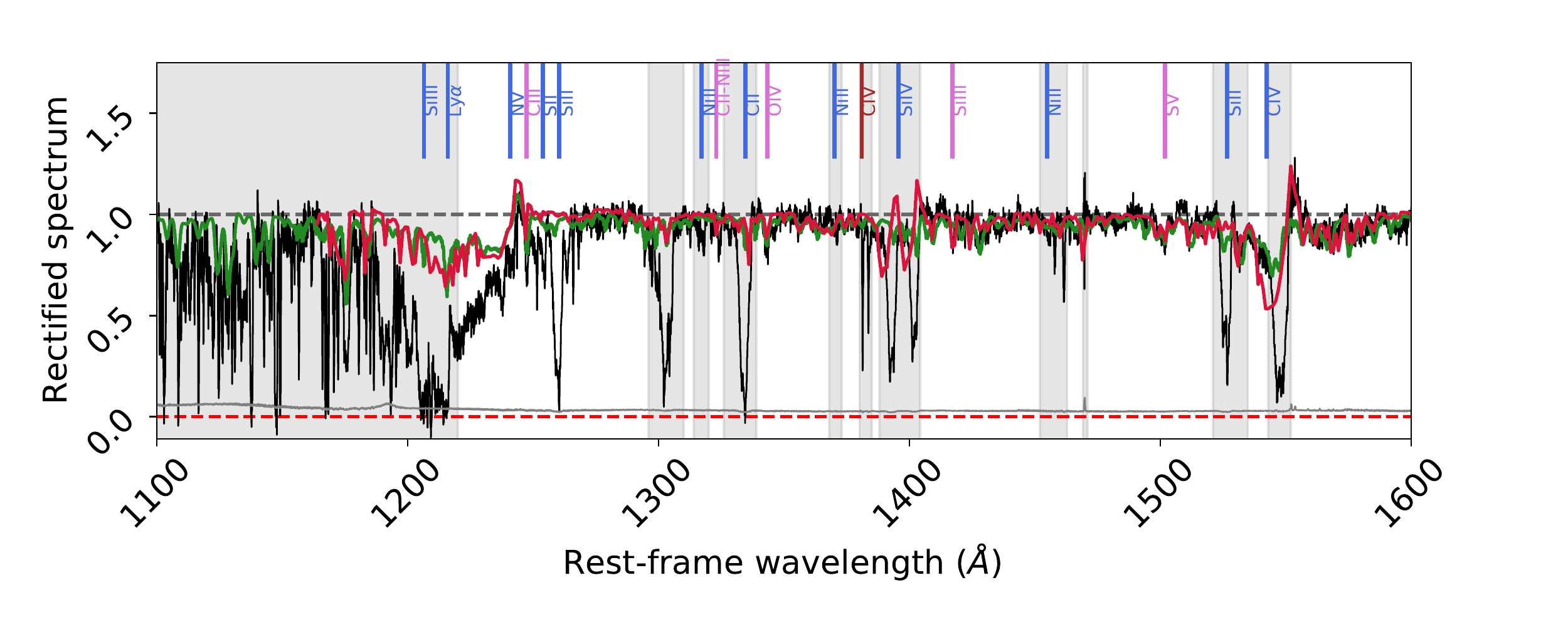}\\
  \includegraphics[width=1\linewidth]{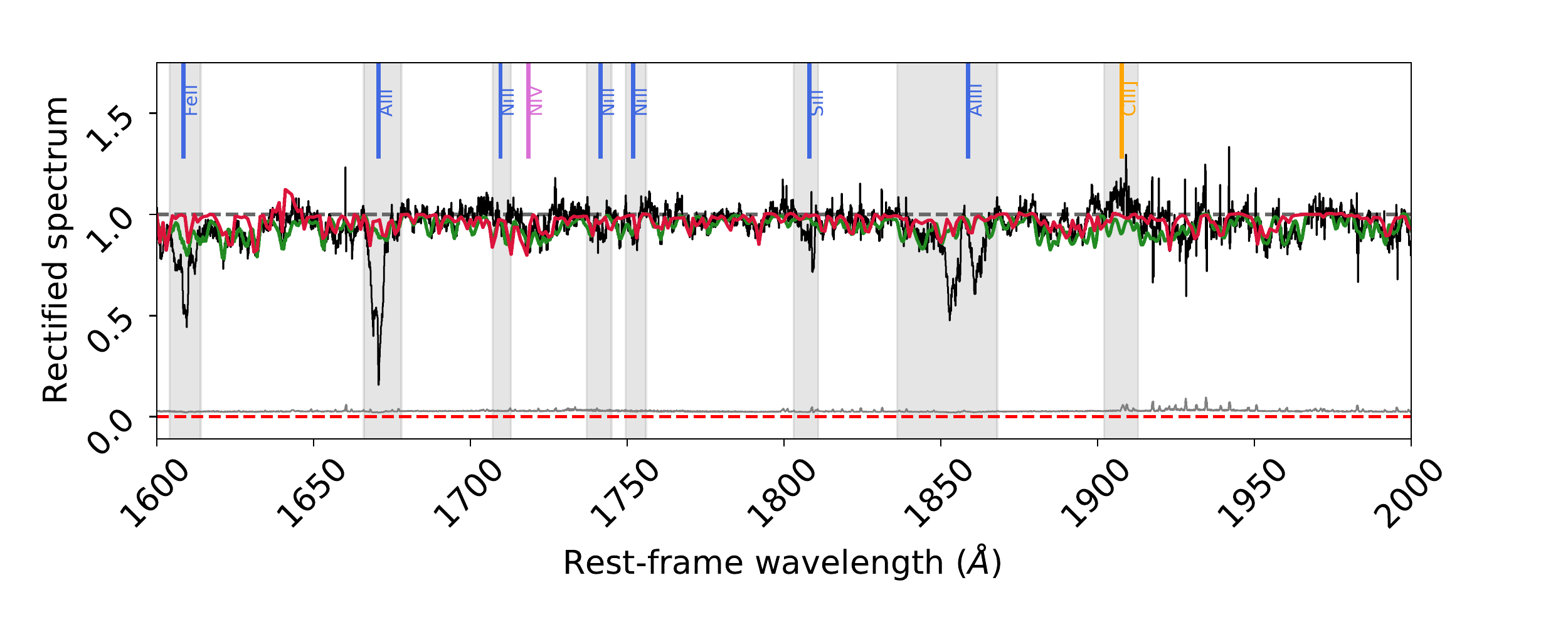}
\caption{Comparison between the J1059 UV spectrum and the best fit {\sc starburst}99 and Binary Population and Spectral Synthesis (BPASS) synthetic spectra \citep{Eldridge+2016} derived using the $\chi^2$ minimization. Grey shaded regions are the wavelength windows excluded from the fit because they include interstellar features. The error spectrum is shown in grey. {The colored vertical lines have the same meaning as in Fig. \ref{fig:ESI_spec}}.}
\label{fig:J1059_fit}
\end{figure*}

The first step in the analysis of this rich spectrum is to establish the systemic redshift of the galaxy. To this end, we use the line list in the UV spectroscopic atlas of starbursts by \cite{Leitherer+2011} to identify photospheric lines which are unblended and sufficiently well defined for their wavelengths to be measured with confidence. We isolate seven suitable photospheric absorption lines, listed in Table~\ref{tab:table_sysz}. We deduce a mean systemic redshift $z_{\rm stars}= 2.79556 \pm 0.00005$ if we take a weighted mean of the entries in Table~\ref{tab:table_sysz}; for comparison the unweighted mean is $z_{\rm stars}= 2.79556 \pm 0.00027$. We adopt the former as our estimate of $z_{\rm sys}$.

\section{Global properties of the stellar population}
\label{sec:general_prop}

In this section we use the broadband photometry of J1059 to constrain its global stellar population properties, in order to place it in context within the high redshift galaxy population.

\subsection{Physical properties from SED fitting}
\label{sec:sedfitting}
%Here we use the \textit{HST} and WISE photometric observations (see Table \ref{tab:table_hst}) corrected for the lensing magnification to model J1059 with SED fitting.We use BPASS v2.2.1 models with stellar metallicity $Z_*=0.002$, a Chabrier IMF, an upper mass limit of 100 $\rm M_{\odot}$, and with binary stellar evolution included. We adopt a constant star formation history and the Small Magellanic Cloud (SMC) extinction curve (\citealp{Lequeux+1982}; \citealp{Prevot+1984}; \citealp{Gordon&Clayton1998}, \citealp{Gordon+2003}), motivated by evidence that the relationship between dust attenuation and the UV slope for $z=1.5$--2.5 galaxies with sub-solar metallicities is consistent with the SMC extinction law \citep{Reddy+2018_2}. 
%We also performed the fit using the \citet{Calzetti+2000} extinction law, but found that this resulted in best-fit models with unphysically young ages.
{Here we use the \textit{HST} and WISE photometry (see Table \ref{tab:table_hst}) corrected for both the lensing magnification and foreground Galactic extinction, for which we find $E(B-V)=0.013$ using the Milky Way dust map of \citet{Green+2015}. We model the J1059 SED using the code Prospector \citep{Johnson+2020}, which employs the Flexible Stellar Population Synthesis models (FSPS;  \citealt{Conroy+2009,Conroy+2010}). We adopt the BPASS v2.2 stellar spectra and libraries, including binary stellar evolution, an upper mass limit of 100 $\rm M_{\odot}$, and a \citet{Chabrier2003} IMF, leaving the metallicity as a free parameter of the fit. We adopt a constant star formation history and the Small Magellanic Cloud (SMC) extinction curve (\citealp{Gordon+2003}),} motivated by evidence that the relationship between dust attenuation and the UV slope for $z=1.5$--2.5 galaxies with sub-solar metallicities is consistent with the SMC extinction law \citep{Reddy+2018_2}. 
We also performed the fit using the \citet{Calzetti+2000} extinction law, but found that this resulted in best-fit models with unphysically young ages ($\rm \lesssim 5\,Myr$), {which are not compatible with the dynamical timescales ($\rm\,\sim 50\,Myr$) inferred for $z\sim2$ galaxies (e.g. \citealp{Erb+2006a}; \citealp{Reddy+2018}).}
%[mass, log(Z/Z_sun), E(B-V), Age(Myr)] = [3.22e+10, -4.63e-01, 5.40e-02, 6.44e+02]

{Markov Chain Monte Carlo (MCMC) modeling of the SED results in a best-fit model with stellar mass $M_*= (3.22\pm 0.20)\,\times10^{10}$ $\rm M_{\odot}$, age $644^{+76}_{-90}$ Myr, $E(B-V)= 0.05\pm 0.012$, and metallicity $Z=0.34^{+0.20}_{-0.13}\, Z_{\odot}$. With a constant star formation history, these results then imply a star formation rate of 50 $\pm$ 7 $\rm M_{\odot}\,yr^{-1}$. We find that while the mass, extinction and age are generally well-constrained, the metallicity probability distribution is significantly broader, with the other parameters equally well-fit for metallicities in the range $Z=0.2$--0.5\,$Z_{\odot}$. The metallicity is better constrained by the spectrum, as we describe in  
Section \ref{sec:stellar_pop}.}

The values derived here for J1059 are typical of galaxies (lensed or not) at $z = 2-3$ (e.g.\ \citealp{Erb+2006},  \citealp{Shapley2011}, \citealp{Santini+2017}, \citealp{Du+2018}, \citealp{Nakajima+2018}, \citealp{Pantoni+2021}).
It is worth pointing out that our SED fitting results can be affected by the well-known degeneracy between age and dust \citep{Papovich+2001}, which have similar effects of reddening the colors of galaxies.
The derived stellar mass and SFR {%(corrected for dust extinction and for the lensing magnification)} are shown in Figure \ref{fig:sfr_mass}, which illustrates J1059 within the SFR-mass plane, together with the main sequences of star forming galaxies at different redshifts derived by \cite{Santini+2017}. 
 of J1059 (corrected for dust extinction and lensing magnification) are consistent with the SF main sequences at $z\sim 2$--3 (e.g. \citealp{Santini+2017}). This suggests that J1059 is not experiencing a short-lived starburst episode (\citealt{Rodighiero+2011}) and can be considered representative of the typical population of star-forming galaxies at high redshift ($z=2-3$).}

\subsection{Star formation rate and UV slope}
\label{sec:dust}
We also estimate the SFR and dust extinction from the ESI spectrum and rest-frame UV photometry alone, for comparison with the results of the photometric SED fitting. We use the \textit{HST} F606W and F814W images, which for the redshift of J1059 have rest-frame central wavelengths 1550 \AA\ and 2120 \AA\ respectively. From these rest-frame UV images only, we find a UV slope of $\beta = -1.37$, where $F_{\lambda} \propto \lambda^\beta$. Restricting the measurement to only the portions of the images falling within the spectroscopic slit results in a slightly bluer slope of $\beta = -1.41$.

We next use the \textit{HST} photometry to finalize the flux calibration of the ESI spectrum, adjusting the spectrum so that synthetic F606W and F814W magnitudes calculated from it match the photometry within the slit. {This calibration is intended to correct for potential wavelength-dependent slit losses due to differential refraction that may cause the spectroscopic and the photometric UV slopes to differ.}
We then fit the calibrated, telluric-corrected spectrum over the rest-frame wavelength range 1270--2300 \AA, with the strong absorption lines masked. The resulting slope is $\beta = -1.61 \pm 0.08$, where the uncertainties are derived from Monte Carlo simulations and come primarily from the flux calibration of the spectrum. This slope is considerably bluer than the value of $\beta$ from the photometry alone we found above; we attribute this difference to the strong absorption lines in the spectrum that fall in the F606W filter, which are masked in the spectral fit but not accounted for in the photometry. We therefore adopt $\beta = -1.61 \pm 0.08$ as the final UV slope. This value is consistent with those found in galaxies at $2<z<5$ in the COSMOS \citep{Taniguchi+2007} and VANDELS \citep{McLure+2018} fields (\citealp{Pilo+2019}, \citealp{Calabro+2020}; see also Section \ref{sec:conc} for further details).

{Adopting the SMC extinction law \citep{Gordon+2003}, we then obtain $E(B-V)=0.06\pm 0.01$, which is in good agreement with the value of $E(B-V)= 0.05\pm 0.012$ found from the SED fitting in Section \ref{sec:sedfitting} above.
We derive an absolute UV magnitude $M_{\rm UV}=-22.63\pm0.02$, roughly two magnitudes brighter than $M^*_{\rm UV}$ at $z\sim2$--3 and placing J1059 at the bright end of the UV luminosity function \citep{Theios+2019}, \citep{ReddySteidel2009}. Using the lensing and dust-corrected F606W magnitude to trace the UV continuum luminosity, we use the \citet{Kennicutt&Evans2012A} relation to obtain an extinction-corrected $\rm SFR = 90\pm7\, M_{\odot}\,yr^{-1}$, somewhat higher than the value obtained by SED fitting. However, we note that \citet{Theios+2019} show that at sub-solar metallicities SFR-UV luminosity calibrations based on the BPASS models result in lower SFRs than the \citet{Kennicutt&Evans2012A} relation. Using their calibration based on BPASS models with a 100 $M_{\odot}$ upper mass cutoff and $Z_*=0.004$ then gives a corrected $\rm SFR = 70 \pm 5\, M_{\odot}\,yr^{-1}$, closer to the value of 50 $\rm M_{\odot}\,yr^{-1}$ obtained from the SED fitting in Section \ref{sec:sedfitting} above. We also recall that the SFR$_{\rm SED}$ is derived from the best-fit stellar mass and age as described above, and therefore depends on the entire SED rather than on the rest-frame UV alone.
Finally, we note that there is an additional $10\%$ uncertainty to be added to the luminosity and stellar mass estimates due to the uncertainty in the magnification correction.}

\section{The stellar spectrum}
\label{sec:stellar_pop}
In this section we analyze the photospheric spectrum of J1059 in order to uncover further its stellar population properties. In particular, we use both the stellar continuum and stellar absorption features to derive the stellar metallicity. In order to perform this study, we fit the observed UV spectrum with a set of {\sc starburst}99 synthetic models \citep{Leitherer+1999}. We adopt 100 Myr old models with a continuous star formation history, which is a reasonable assumption for galaxies which are undergoing current star formation. Moreover, an age of 100 Myr ensures that the synthetic UV spectra are stable against the fast evolution of very massive stars, which occurs on timescales shorter than $\sim$ 30 Myr. We adopt a Salpeter initial mass function (IMF) \citep{Salpeter1955} with a upper mass limit of 100 M$_\odot$ (note that the Salpeter and the Chabrier IMF are the same in the high mass regime). We consider five different stellar metallicities: $Z_*=0.001, 0.004, 0.008, 0.02$ and  $0.04$, where $Z_*=0.02$ is the solar value (in these models). We assume Geneva evolutionary tracks with high mass loss \citep{Meynet+1994}, which are able to model the stellar winds and thus the P-Cygni profiles characterizing some of the absorption features within the considered wavelength range. 

Before comparing the {\sc starburst}99 model spectra with the data by means of a $\chi^{2}$ minimization, we smooth the former to match the velocity dispersion of the stars in J1059 as recorded with ESI. From the widths of the stellar lines listed in Table~\ref{tab:table_sysz}, we measure a velocity dispersion $\sigma = 130$\,km~s$^{-1}$, {a value typical of rather massive ($M_*\sim 10^{10}\,$M$_{\odot}$) star-forming galaxies at $z = 2$--3} \citep[see, for example,][]{Erb+2006, Forster_Schreiber+2006}. We therefore smooth the {\sc starburst}99 output spectra accordingly and rebin them to the 20\,km~s$^{-1}$ wide bins of the ESI spectrum. Since {\sc starburst}99 is designed to model only the stellar component of a galaxy spectrum, we mask out the interstellar and nebular features before carrying out the comparison. We define the $\chi^{2}$ as:

\begin{equation}
   \rm \chi^2=\sum_{i}{(O_{\lambda}-M_{\lambda})^2/e_{\lambda}^2}\, ,
    \label{eq:eq_chi2}
\end{equation}
where $\rm O_{\lambda}$ is the observed spectrum, $\rm M_{\lambda}$ is the model considered for the fit and $\rm e_{\lambda}$ is the error on the observed spectrum.

\begin{figure*}
	\includegraphics[width=0.95\linewidth]{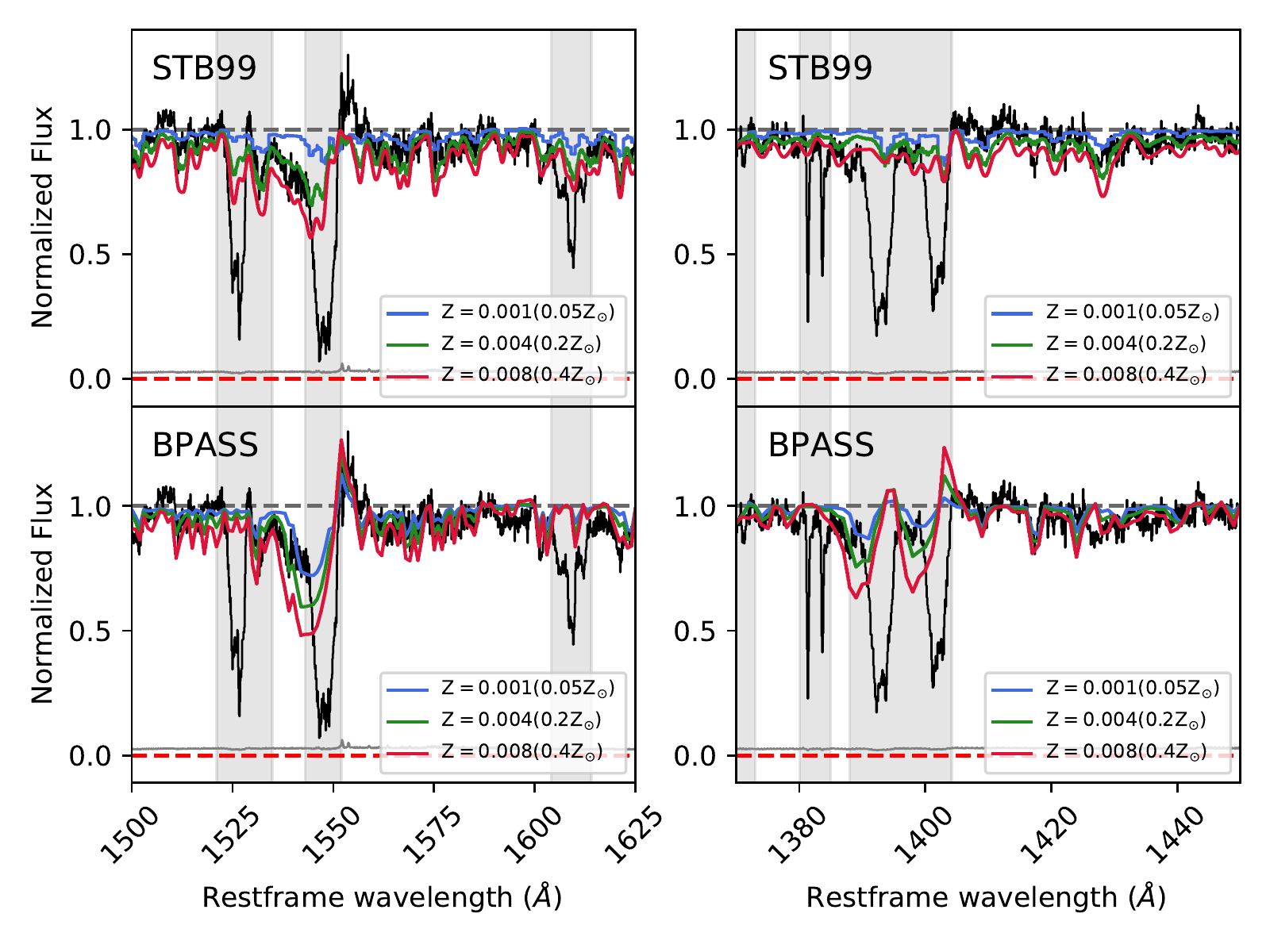}
	 \caption{Comparison between the ESI spectrum of J1059 in the region encompassing the C IV and the Si IV lines (black) and the synthetic spectra described in the text. The top panels show three {\sc starburst99} models with $Z=0.001$,$Z=0.004$ (best fit) and $Z=0.008$; the bottom panels show three BPASS models (including the effects of binary evolution) with $Z=0.001$, $Z=0.004$ (best fit) and $Z=0.008$. Grey vertical shaded regions corresponds to the wavelength ranges that have been excluded from the fit.}
   \label{fig:J1059_fit_civ}
\end{figure*}

By performing the $\chi^2$ minimization, we derive a stellar metallicity $Z_*=0.004^{+0.0045}_{-0.001}$ (between $\sim 0.15$ and $\sim 0.5\, Z_\odot$). The errors represent the 68\% confidence interval, obtained from 100 Monte Carlo re-simulations of the ESI spectrum. Figure \ref{fig:J1059_fit} shows the J1059 normalized spectrum together with the best fit {\sc starburst}99 model. It is possible to appreciate that the model is able to reproduce the multitude of low contrast photospheric features in the UV spectrum over the whole wavelength range \citep{Leitherer+2011}.
The ability of the fit to reproduce the observed spectrum also suggests that continuous mode models and a Salpeter IMF including stars as massive as 100 $M_{\odot}$ give a good representation of the properties of J1059. 

We also fit the observed spectrum with BPASS models \citep{Eldridge+2016}, which include massive binary stars in the stellar population. We adopt BPASS models with properties as similar as possible to the {\sc starburst}99 ones, i.e.\ with a 100 Myr old continuous star formation history, metallicities in the range $Z=0.001-0.04$ and a Salpeter IMF up to 100 $M_{\odot}$. We find that the BPASS best fit metallicity is $Z_*=0.004^{+0.006}_{-0.001}$ (68\% confidence interval, as before), in broad agreement with the {\sc starburst}99 results.

The stellar mass and stellar metallicity we derived for J1059 fall on the stellar mass-stellar metallicity relation found by \citet{Cullen+2019} through UV spectral fits of VANDELS \citep{McLure+2018} galaxies with $M_*>10^{10}\,M_{\odot}$ and in the redshift range $2<z<5$. They are also consistent with the stellar mass-stellar metallicity relation derived by \citet{Calabro+2020} by means of the photospheric absorption indices defined by \citet{Rix+2004}. 
The observed metallicity suggests rapid star formation that has polluted the ISM in a relatively short amount of time ($\sim$ 500 Myr), giving birth to metal-enriched generations of stars. The general stellar properties of J1059 confirm the trends and the properties found so far that characterize the population of high-$z$ galaxies. Moreover, they show how different methods (i.e.\ full-UV spectrum fitting and photospheric absorption indices) are consistent in predicting the stellar properties of high redshift galaxies (see Section \ref{sec:conc} for further details).

Figure \ref{fig:J1059_fit_civ} shows the portions of the J1059 spectrum encompassing the C {\sc IV} $\lambda\lambda$1548, 1550 and the Si {\sc IV} $\lambda\lambda$1393, 1402 absorption features, compared to three {\sc starburst}99 and three BPASS synthetic spectra with different metallicities. The C {\sc IV} $\lambda\lambda$1548, 1550 and Si {\sc IV} $\lambda\lambda$1393, 1402 lines include both a stellar and an interstellar component, which can be easily distinguished from one another thanks to the high resolution of our data. These lines show the typical P-Cygni profile arising from stellar winds and consisting of a blueshifted absorption component and a redshifted emission component. 
The Si {\sc IV} P-Cygni stellar feature is stronger in more evolved and/or higher metallicity stellar populations, where Si$ ^{+++}$ is the dominant ionized species (see \citealp{Chisholm+2019} for further details), so it may be unsurprising that J1059 does not show noticeable Si {\sc IV} P-Cygni emission.

As can be seen in the figure, the $Z_*=0.02\,Z_{\odot}$ best fit {\sc starburst}99 model is able to reproduce the \textit{absorption} portion of the P-Cygni profile of the C {\sc IV} line, while the lower ($Z_*=0.05\,Z_{\odot}$) and the higher ($Z_*=0.04\,Z_{\odot}$) metallicity models underpredict and overpredict it, respectively. However, the best fit model itself underpredicts the \textit{emission} portion of the P-Cygni profile. Looking at the BPASS models, they are able to reproduce the \textit{emission} portion of the P-Cygni profile, although overestimating the \textit{absorption}. The situation is reversed for Si {\sc IV}, where the {\sc starburst}99 models do a better job in reproducing the whole P-Cygni profile than the BPASS models.

\begin{figure*}
	\includegraphics[width=\linewidth]{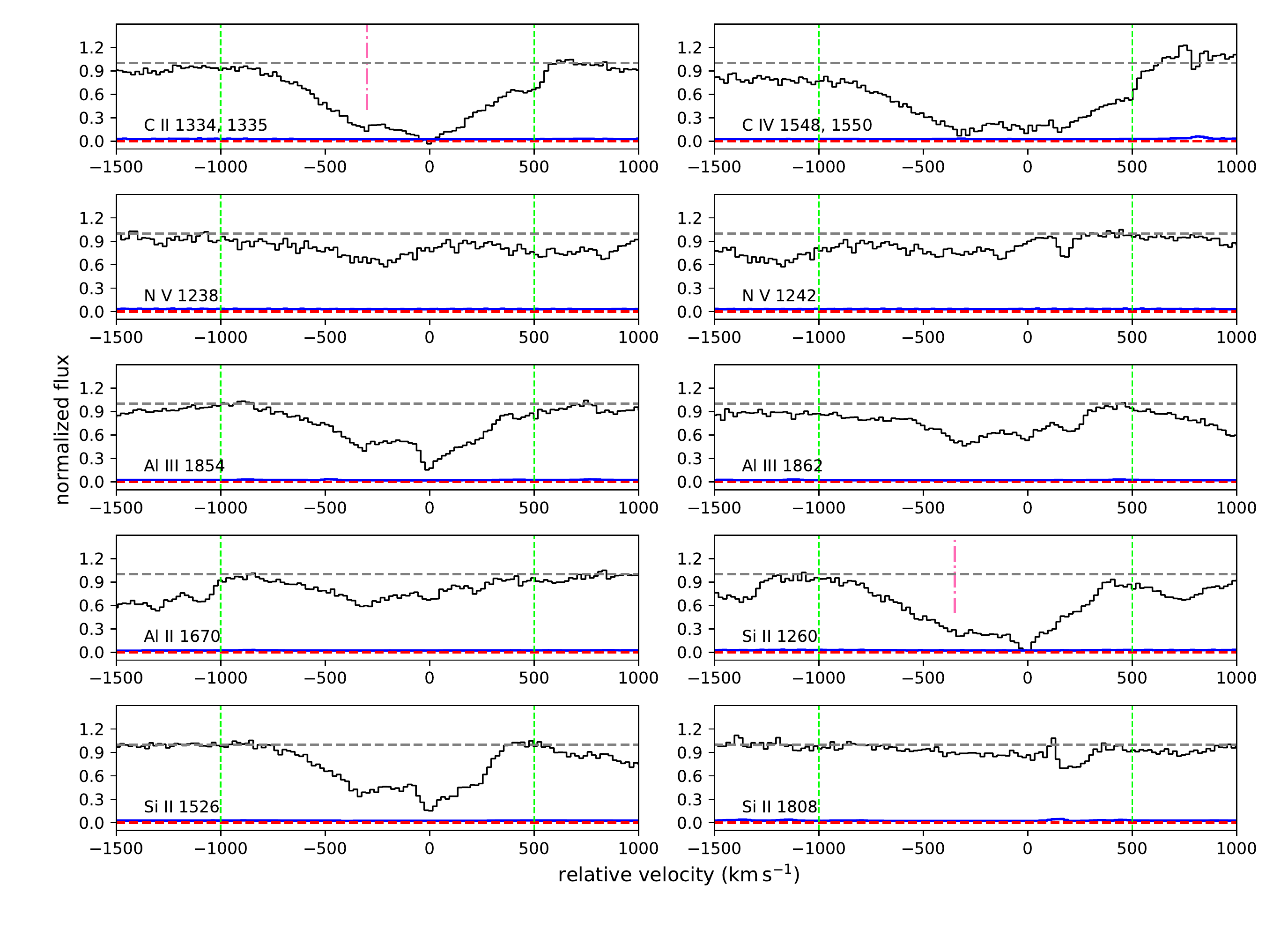}
       \caption{Velocity profiles of interstellar absorption lines. Velocities are relative to the systemic redshift $z_{\rm stars}=2.79556$. The blue line (just visible above the red dash line indicating the zero level) is the error spectrum. The two vertical green dash lines indicate the limits of integration for the measurements of equivalent widths and column densities (see Sections~\ref{sec:kinematics} and \ref{sec:MetalLines}). Pink vertical lines show the blending of some of the lines with other spectral features (see Table \ref{tab:table_na}). We also remind that C {\sc IV} 1548, 1550 are blended with each other.}
   \label{fig:ISM_lines_1}
\end{figure*}

A possible explanation of the observed C {\sc IV} emission might be the presence of redshifted nebular emission near 1551.6 \AA\, \citep{Leitherer+2002}. This is not unusual and has already been found in other high redshift lensed galaxies (e.g.\ the Cosmic Horseshoe and the Cosmic Eye; \citealt{Quider+2009}, \citeyear{Quider+2010}). There are also known cases, among strongly lensed high redshift galaxies, where nebular emission is the dominant component of the C {\sc IV} feature (see \citealp{Christensen+2012}, \citealp{Stark+2014}, \citealp{Smit+2017}, \citealp{Berg+2018}). However, 
the ability of the BPASS models to reproduce that same emission suggests that it might have a stellar origin, probably related to Wolf-Rayet (WR) stars. This latter hypothesis would imply that the BPASS models provide a better treatment of the WR stars and their evolution than {\sc starburst}99. On the other hand, Si {\sc IV} does not show a strong P-Cygni emission component, and the BPASS models that reproduce the C {\sc IV} emission feature overpredict the Si {\sc IV} emission.

{The discrepancies between the {\sc starburst}99 and BPASS models are probably related to a different treatment of mass loss, stellar rotation and stellar multiplicity. On the one hand, the BPASS models include binary evolution, which increases the lifetime of more massive stars producing hotter, bluer stellar spectra. On the other hand, the {\sc starburst}99 models we are using do not include stellar rotation and therefore predict a lower mass loss rate, which weakens the stellar features associated with winds. These discrepancies show that further improvements are needed even in state-of-the-art stellar synthetic spectra to theoretically reproduce all the variables involved in stellar evolution.}

{A further potential complication is that the {\sc starburst}99 and BPASS models do not consider non-solar element abundance patterns, such as the overabundance of alpha-capture elements relative to Fe-peak elements that seems to be a common feature of galaxies undergoing rapid star formation (e.g. \citealp{Steidel+2016}, \citealp{Strom+2018}), and is indeed
 found to apply to the interstellar gas in J1059 (see Section~\ref{sec:chemical_comp} and Figure~\ref{fig:fig_abd_1}.) 
This is a shortcoming of the synthetic models that has not yet been addressed. 
The 1100--2000\,\AA\ wavelength range covered by our ESI spectrum includes many spectral features from a variety of elements of the periodic table, from H to Ni. Consequently, the stellar metallicity we have derived here has to be considered as an approximate measure of the average degree of metal enrichment achieved by the early-type stars in J1059.}
%This is expected, since the Si {\sc IV} P-Cygni stellar feature is stronger in more evolved and/or higher metallicity stellar populations, where Si$ ^{+++}$ is the dominant ionized species (see \citealp{Chisholm+2019} for further details). The strong Si {\sc IV} absorption we see is therefore mostly of interstellar origin.}

\begin{figure*}
	\includegraphics[width=0.95\linewidth]{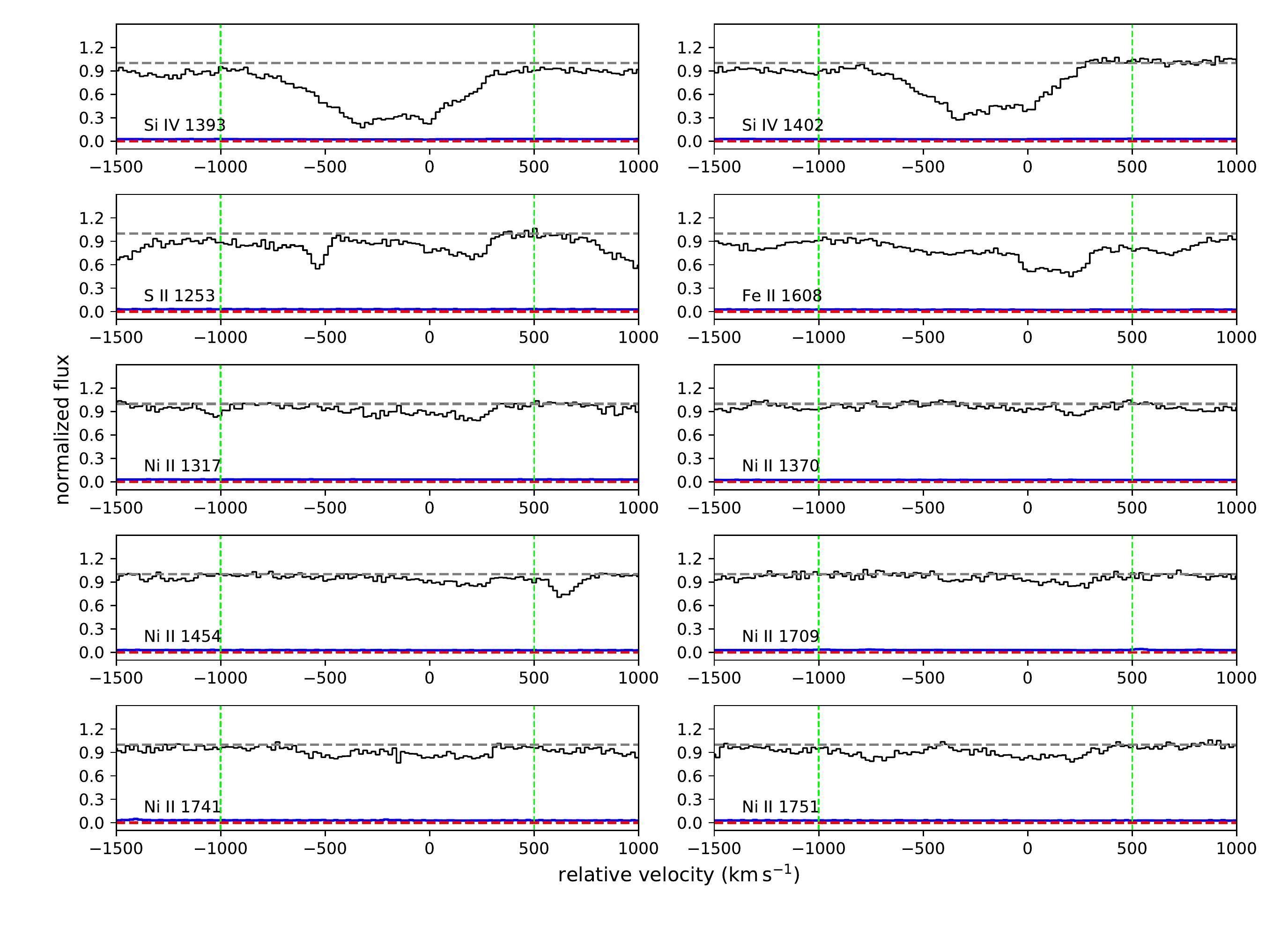}
      \caption{Velocity profiles of interstellar absorption lines. Velocities are relative to the systemic redshift $z_{\rm stars}=2.79556$. The blue line (just visible above the red dash line indicating the zero level) is the error spectrum. The two vertical green dash lines indicate the limits of integration for the measurements of equivalent widths and column densities (see Sections~\ref{sec:kinematics} and \ref{sec:MetalLines}).}
   \label{fig:ISM_lines_2}
\end{figure*}

\section{The Interstellar Spectrum}
\label{sec:ISM}

We now turn to the analysis of the interstellar spectrum of J1059, with its attendant information on the kinematics and chemical composition of diffuse gas in the galaxy.

\subsection{Gas Kinematics}
\label{sec:kinematics}

We identify 20 interstellar absorption lines (or blends of absorption lines) in the ESI spectrum of J1059, produced by elements from H to Ni, in a range of ionization stages from neutral hydrogen to four-times ionized nitrogen. The profiles of all 20 absorption lines are plotted in Figures~\ref{fig:ISM_lines_1} and \ref{fig:ISM_lines_2} on a velocity scale relative to the systemic redshift $z_{\rm stars} = 2.79556$. These plots show clearly the large velocity extent of the lines:\ there is gas at both negative and positive velocities relative to the redshift of the stars.

\begin{figure*}
	\includegraphics[width=0.95\linewidth]{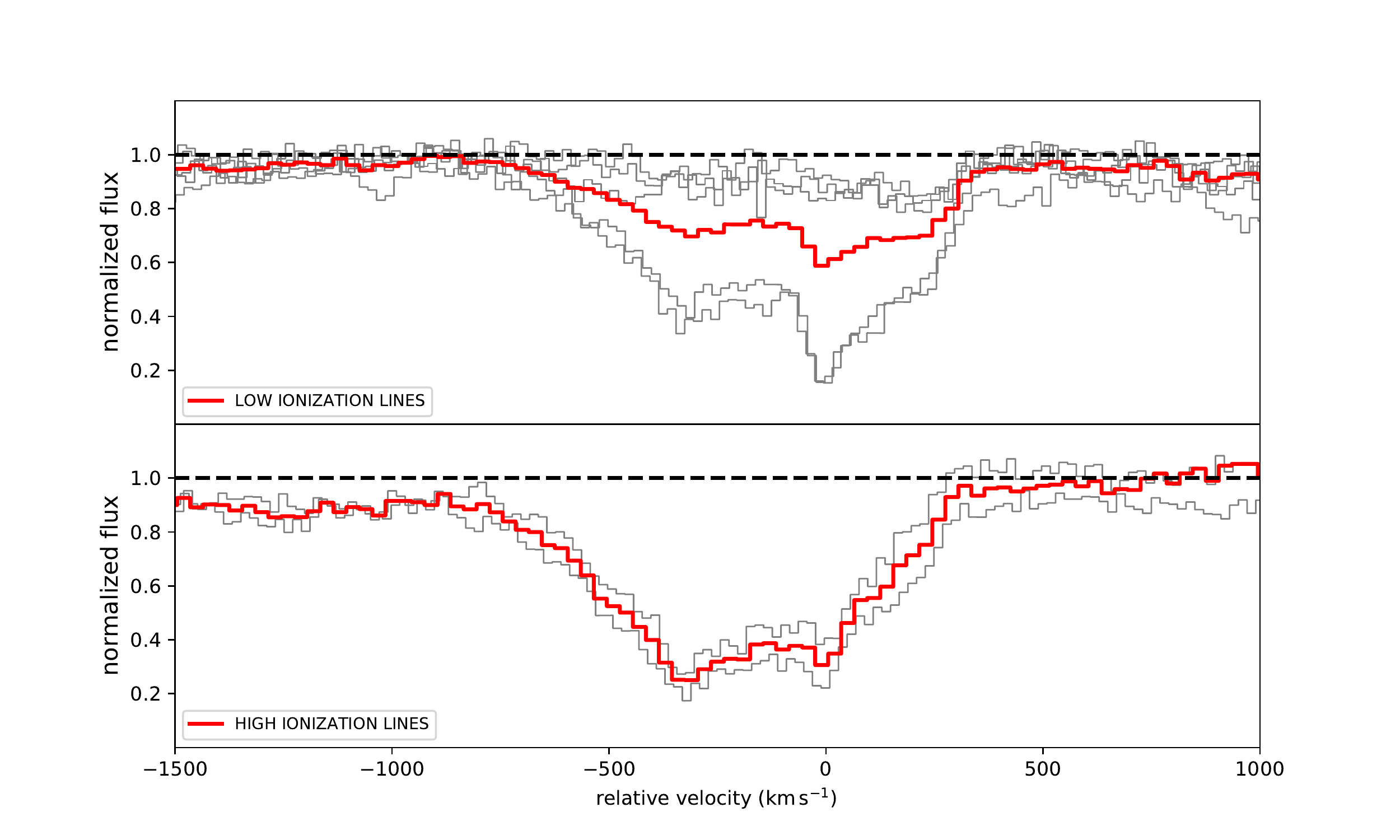}
      \caption{Average velocity profiles for low and high ionization lines. Grey curves are the individual ISM lines considered to derive the average, the red curves are the average profile.}
   \label{fig:ISM_lines_avg}
\end{figure*}

To better assess the full velocity extent of the absorbing gas, we construct average line profiles using all of the absorption lines, or portions of lines, that are not blended; we do this separately for species that are the dominant ionization stages of their elements in neutral gas (i.e.\ the first ions for the species considered here), and for more highly ionized species. In particular, for the low ionization lines, we only considered the Ni {\sc ii} $\lambda$1709, Ni {\sc ii } $\lambda$1741, Si {\sc ii} $\lambda$1526 and Al {\sc ii} $\lambda$1670 lines. For the high ionization, we only considered the Si {\sc iv} $\lambda$1393 and Si {\sc iv} $\lambda$1402 lines. The average profiles, reproduced in Figure \ref{fig:ISM_lines_avg}, show absorption extending over the range $v \simeq -800$ to $\simeq +300$\,km~s$^{-1}$ for both neutral and ionized gas, although the balance between blueshifted and redshifted absorption is somewhat different between the two. 

Blueshifted interstellar absorption is a common feature of star-forming galaxies at high as well as low redshifts \citep[e.g.][]{Heckman+2000, Pettini+2001, Shapley+2003, Steidel+2010, Marques-Chaves+2020} and is generally interpreted as tracing galaxy-wide outflows powered by the kinetic energy and momentum deposited into the ISM by the starburst \citep[see][for comprehensive reviews]{Veilleux+2005, Heckman+2015, Veilleux+2020}. That the outflowing gas should reach velocities as high as nearly $\sim -1000$\,km~s$^{-1}$ is not unexpected either; such an extended blue wing to the line profiles has been recorded in other well-studied gravitationally lensed galaxies at $z = 2$--3 \citep[e.g.][]{Pettini+2002, Quider+2009, Dessauges-Zavadsky+2010}. 

{On the other hand, in J1059 the absorption at positive velocities relative to the stars is somewhat stronger and extends further (kinematically) than in previous well-observed examples. If the interstellar medium of this galaxy has a similar velocity dispersion as the stars
($\sigma = 130$\,km~s$^{-1}$; see Section~\ref{sec:stellar_pop}), some of this redshifted absorption (as well as some of the blue wing)
could simply be the high velocity tail of the ambient ISM. The profiles of the high ionization absorption lines (see lower panel of
Figure~\ref{fig:ISM_lines_avg}) are consistent with this interpretation. However, this is not entirely the case with the low ionization lines,
which evidently show a discrete absorption component, rather than a smooth wing, at positive velocities.
}

We examined the integral field data of James et al.\ (in preparation) to check if the redshifted and blueshifted absorption could be ascribed separately to the two regions making up the source (see right panel of Figure \ref{fig:J1059_LensModel}) and found that this is \textit{not} the case, although there are some differences in the velocity profiles presented by gas in front of each region. Presumably, much of the absorption we see arises in material well in front of the stars and covering both concentrations of stellar light. In a major new study of more than 200,000 foreground-background galaxy pairs at $z\sim2$, \citet{Chen+2020} found that galaxy-scale outflows dominate the kinematics of the gas in the circumgalactic medium (CGM) out to distances of $\sim 50$\,pkpc from the starburst, while at a distance beyond $\sim 100$\,pkpc infall of accreting gas takes over. Within this picture, we may be viewing J1059 along a line of sight that captures a cold filament of dense (given the strength of the absorption lines) accreting material \citep{Dekel+2009}, as well as the near-ubiquitous galaxy-wide outflow.

In columns 4 and 5 of Table~\ref{tab:table_na} we list the values of equivalent width (EW) and associated errors ($\delta$EW) for the 20 interstellar absorption lines in J1059. EWs are measured by integrating the velocity profiles shown in Figs.~\ref{fig:ISM_lines_1} and \ref{fig:ISM_lines_2} over the range $v = -1000$ to $+500$\,km~s$^{-1}$. The errors are calculated with a Monte Carlo approach:\ we perturb the observed spectrum in the region of each line by a random realization of the noise spectrum, and then refit the continuum and recalculate the EW. We repeat this procedure 500 times for each line; the resulting distributions of values of EW are found to be approximately Gaussian in shape, and we adopt their dispersion ($\pm 1 \sigma$) as a measure of the error in EW.

\begin{table*}
	\centering
	\caption{Interstellar absorption lines.}
	\label{tab:table_na}
	\begin{threeparttable}
	\begin{tabular}{lllccccc} % four columns, alignment for each
		\hline
		Ion & ~~~~~$\lambda^{\tnote{a}}$ & ~~~${f}^{\tnote{a}}$ & $\rm EW$ & $\rm \delta\,EW$ & $\log\,(N/\rm cm^{-2})$ & $\log (\delta N/\rm cm^{-2})$\\
		
			 & ~~~~\,$(\si{\angstrom})$ & & $~(\si{\angstrom})$& $~(\si{\angstrom})$ &  &  \\
		\hline

C\,{\sc ii}& 1334.532$^{\tnote{b}}$   & 0.129 & 3.63 & $\pm\,0.16$ & >15.54 &\\
C\,{\sc iv} & 1548.202, 1550.774$^{\tnote{c, e}}$ & 0.2848  & 4.75& & \\
N\,{\sc v} &1238.821$^{\tnote{c}}$ & 0.156 &  2.04&$\pm\,0.10$ & 15.08 & $\pm\,0.02$\\
N\,{\sc v} &1242.804$^{\tnote{c}}$ & 0.0777  &  1.12&$\pm \,0.08$ & 15.09 & $\pm\,0.03$\\
Al\,{\sc ii} & 1670.7867$^{\tnote{f}}$ & 1.74 & 2.79 & $\pm\,0.13$ &13.97 & $\pm\,0.02$\\
Al\,{\sc iii}& 1854.7164$^{\tnote{c,f}}$& 0.561 &   2.36& $\pm\,0.11$ &14.23 &  $\pm\,0.02$\\
Al\,{\sc iii} &1862.7895& 0.279 & 1.62& $\pm\, 0.08$ & 14.34 & $\pm\,0.02$\\
Si\,{\sc ii} &1260.42$^{\tnote{d}}$ & 1.20 & 3.14& $\pm\,0.14$ & >14.53 & \\ %$\pm\,0.10$\\
Si\,{\sc ii}& 1526.72~~ & 0.144~~ & 2.36& $\pm\,0.11$  & >15.08 & \\ %$\pm\,0.03$\\
Si\,{\sc ii}& 1808.00 & 0.00245  &  0.74 & $\pm\,0.07$ &  16.06 & $\pm\,0.04$\\
Si\,{\sc iv} & 1393.76 & 0.513  & 2.83& $\pm\,0.13$ & 14.70 & $\pm\,0.02$\\
Si\,{\sc iv}& 1402.77 & 0.254  & 1.99& $\pm\,0.09$ & 14.81 &  $\pm\,0.02$\\
S\,{\sc ii} &1253.805& 0.0104 &  1.31 & $\pm\,0.08$ & 16.02 &  $\pm\,0.03$\\
Fe\,{\sc ii}& 1608.45078$^{\tnote{f}}$ & 0.0591  &  1.99& $\pm\,0.1$ & 15.26 & $\pm\,0.02$\\
Ni\,{\sc ii} & 1317.217  & 0.0818 & 0.53& $\pm\,0.04$ & 14.65& $\pm\,0.04$ \\
Ni\,{\sc ii} &  1370.132 & 0.0811 & 0.21& $\pm\,0.05$ & 14.21 & $\pm\,0.14$ \\
Ni\,{\sc ii} & 1454.842& 0.0347 &0.28& $\pm\,0.04$ &14.65 & $\pm\,0.05$\\
Ni\,{\sc ii} & 1709.604 & 0.0551  &  0.24 & $\pm\,0.06$ & 14.27 & $\pm\,0.12$\\
Ni\,{\sc ii} & 1741.553 & 0.0488  & 0.69 &  $\pm\,0.07$ & 14.75 & $\pm\,0.05$\\
Ni\,{\sc ii} & 1751.910 & 0.0361  & 0.80& $\pm\,0.07$ &14.94 & $\pm\,0.04$\\
		\hline
	\end{tabular}
	\begin{tablenotes}
  \item[a] Rest wavelengths and $f$-values from \cite{Cashman+2017}.
  \item[b] Blended with C\,{\sc ii}$^\ast$ $\lambda 1335.7077$
  \item[c] Partially blended with each other
  \item[d] Blended with S\,{\sc ii} $\lambda 1259.519$
  \item[e] Blended with stellar P-Cygni line (emission + absorption)
   \item[f] Mildly saturated.
\end{tablenotes}
   \end{threeparttable}
\end{table*}

\begin{figure*}
	\includegraphics[width=\columnwidth]{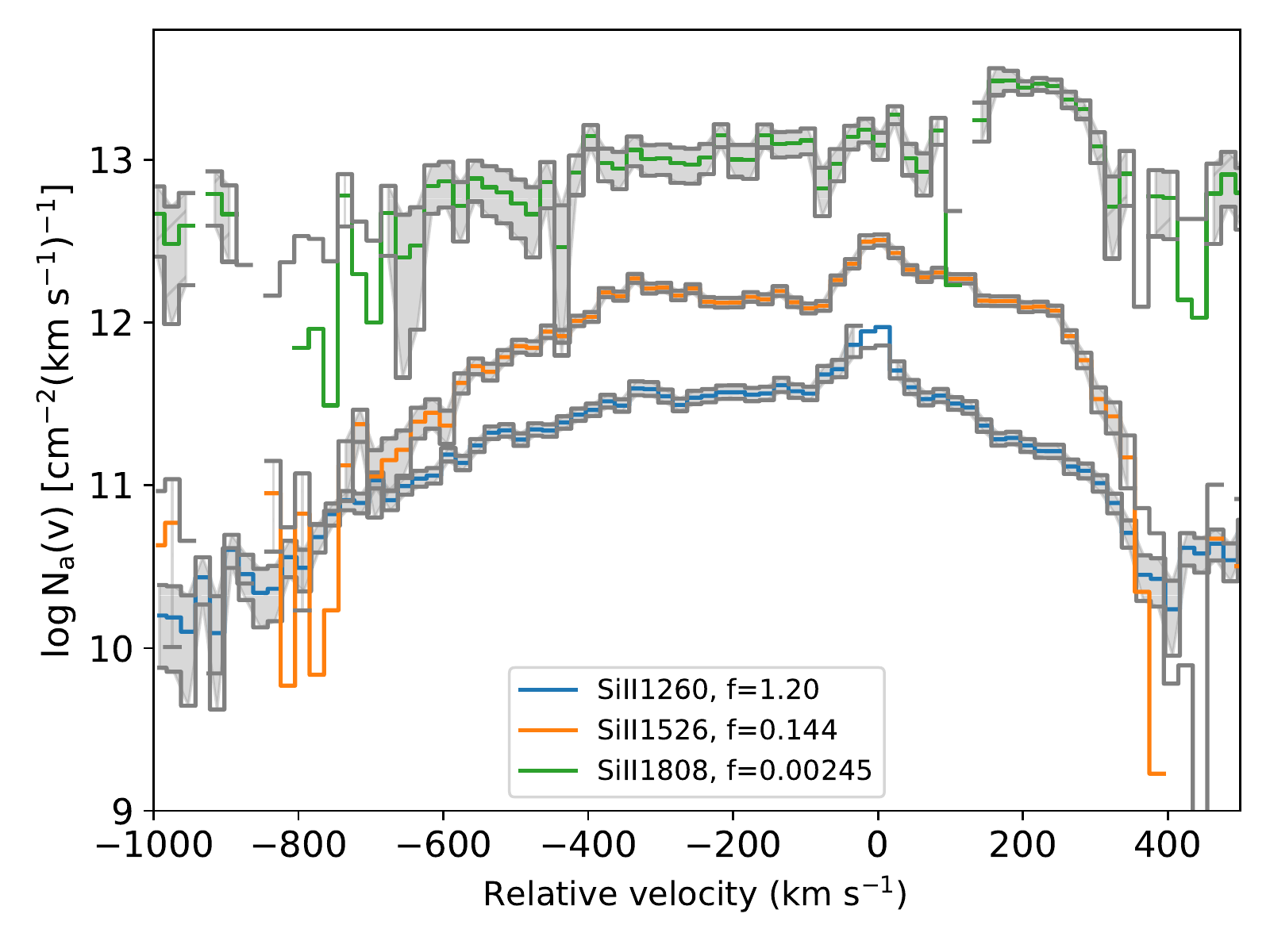}~~~~~\includegraphics[width=\columnwidth]{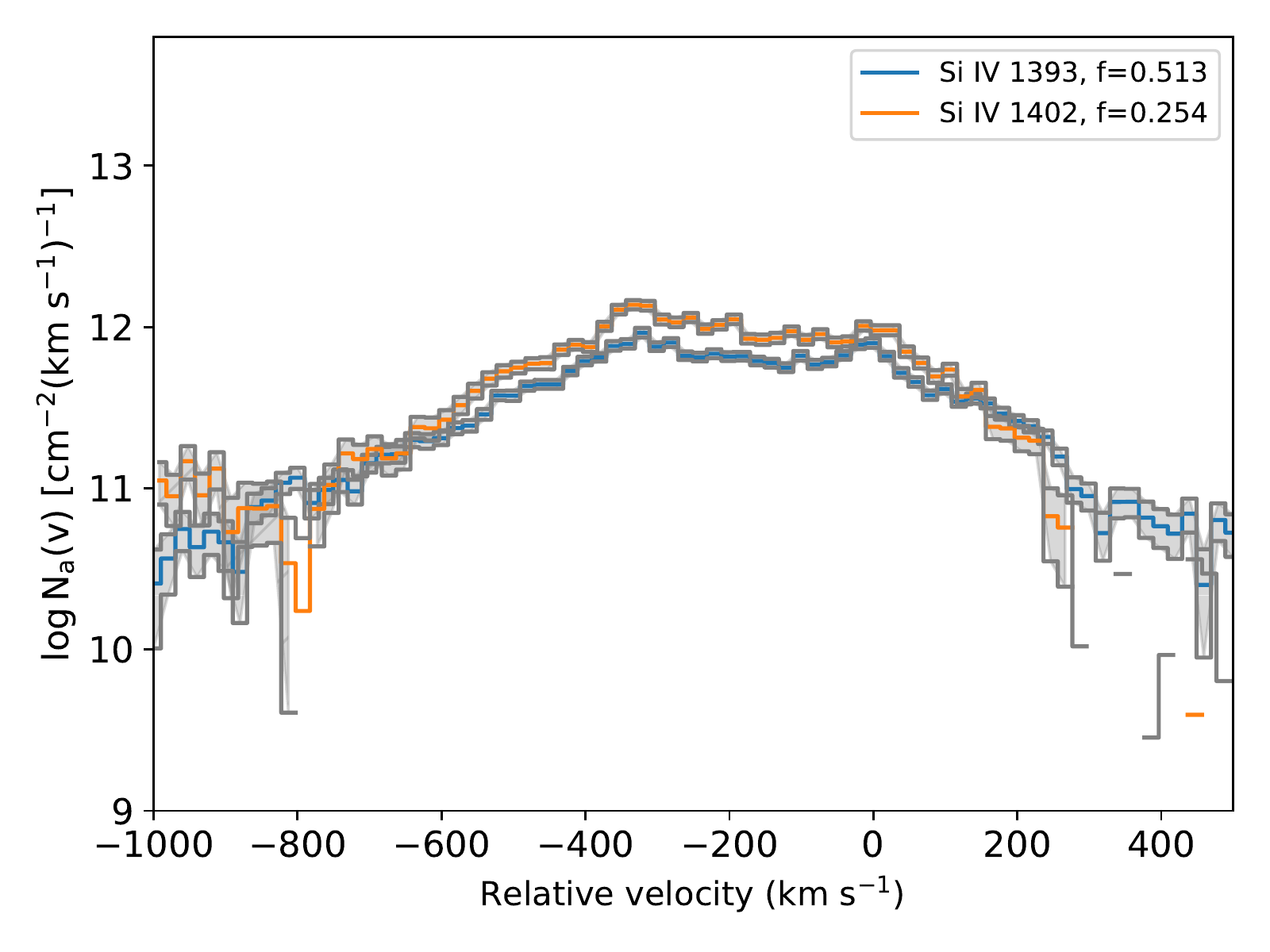}
      \caption{$N_{\rm a}(v)$ profiles for three Si\,{\sc ii} lines (left) and the Si\,{\sc iv} doublet (right). Different colors indicate transition of the same ion but with different oscillator strengths. The grey shaded areas mark the error on the apparent column density.}
   \label{fig:fig_nav_1}
\end{figure*}

\section{Column Densities}
\subsection{Metal Lines}
\label{sec:MetalLines}

The resolved profiles of the interstellar absorption lines recorded with ESI allow us to use the apparent optical depth (AOD) method of \cite{SavageSembach1991} to deduce the ion column densities from unsaturated absorption lines. As those authors pointed out, the advantage of the AOD approach is that it highlights cases where narrow, saturated components are masked by overlapping broader ones; this is a distinct possibility for our data, given the great velocity extent of the lines and the knowledge that the interstellar absorption we see is a composite of many unresolved sightlines with potentially widely differing optical depths.

In the AOD method, the column density of an ion per velocity bin, $N_{\rm a}(v)$ (cm$^{-2}$), is related to the apparent optical depth in that bin, $\tau_{\rm a}(v)$, by the expression:
\begin{equation}
    N_{\rm a}(v)=3.768\,\times\,10^{14}\frac{\tau_{\rm a}(v)}{\lambda\,\textit{f}},
\label{eq:AOD}
\end{equation}
where $\lambda$ and \textit{f} are, respectively, the wavelength (in \AA) and oscillator strength of the atomic transition. Hidden saturation is revealed by discordant values of $N_{a}(v)$ returned from lines with differing $f$-values absorbing from the same ground state of an ion (since by definition there is only one value of the column density for that ground state). Partial, as opposed to complete, coverage of the stars by the absorbing gas would produce a similar effect; however, in our case we do not expect this to be a significant complication, as the cores of the strongest lines in Figure \ref{fig:ISM_lines_1} reach down to the zero level. We calculate values of $N$ for each absorption line by integrating Equation~\ref{eq:AOD} from $-1000$ to $+500$\,km~s$^{-1}$; the associated errors, $\delta N$, are estimated with the same Monte Carlo approach used for the errors in the equivalent widths. Values of $\log N$ and $\log \delta N$ are listed in the last two columns of Table~\ref{tab:table_na}.

\begin{figure*}
   \includegraphics[width=0.95\linewidth]{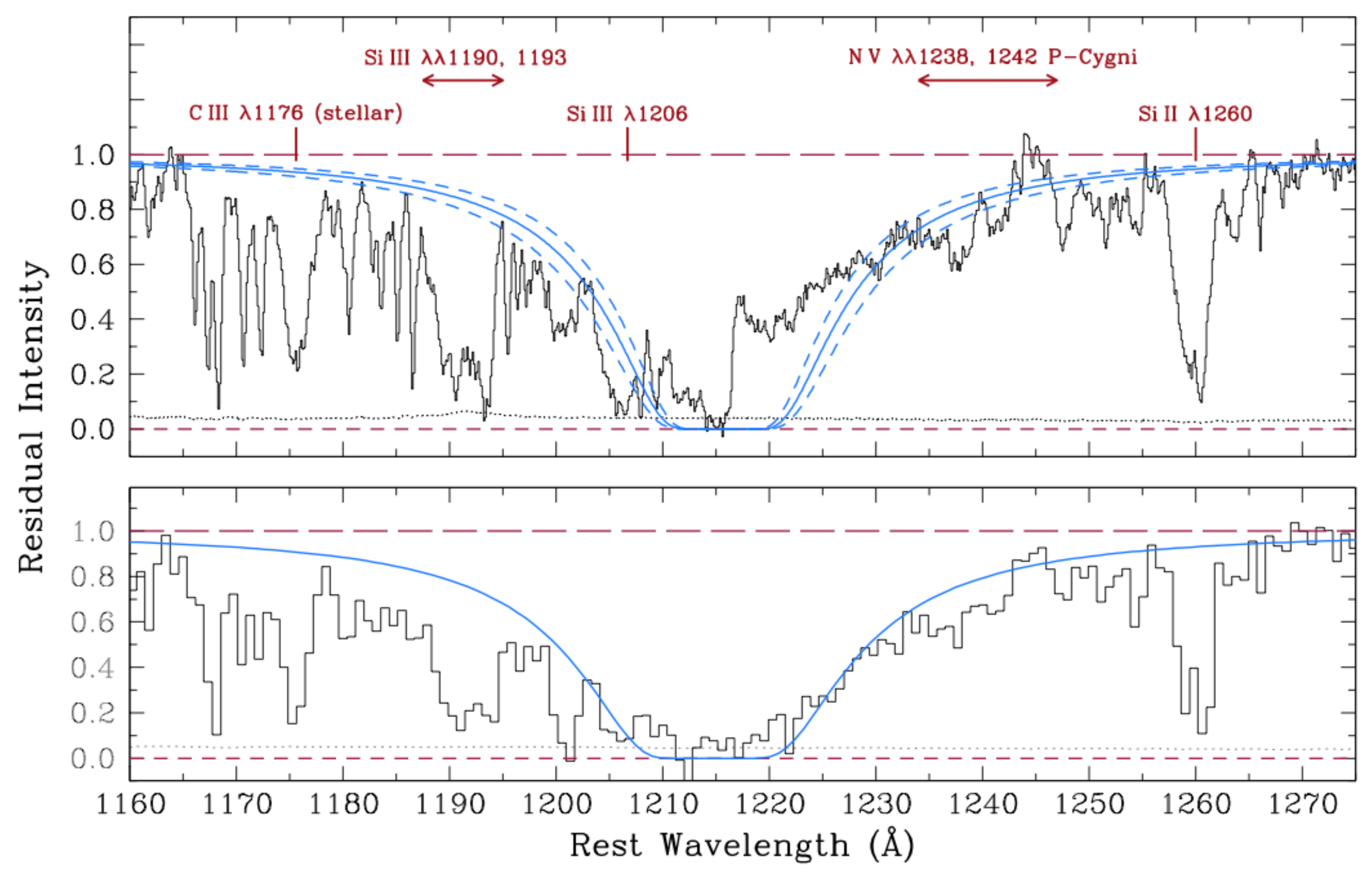}
   \caption{\textit{Upper panel}: Portion of the ESI spectrum (and its $1 \sigma$ error) of J1059 in the region of the Ly$\alpha$ line (black histogram and dotted line respectively). The blue continuous line is a damped profile with column density [$\log N$(H\,{\sc i})/cm$^{-2}] = 21.4$, while the two dash blue lines shows the uncertainty of $\pm 0.1$\,dex in $\log N$(H\,{\sc i}). The Ly$\alpha$ in J1059 is a complex blend of emission and damped absorption; its interpretation is further complicated by overlapping interstellar and stellar features (in addition to the ubiquitous Ly$\alpha$ forest). The strongest of these features are labelled.
   \textit{Lower panel}: The same portion of the J1059 spectrum extracted from the KCWI cube at the location within the arc where the Ly$\alpha$  is (almost) pure absorption. The continuous blue line shows a damped profile with column density [$\log N$(H\,{\sc i})/cm$^{-2}] = 21.54$. See Section~\ref{sec:Lya} for further details.}
\label{fig:lya}
\end{figure*}

In Figure \ref{fig:fig_nav_1} we reproduce two examples of the run of apparent optical depth with velocity. The three Si\,{\sc ii} transitions covered by our data have widely different $f$-values (see Table~\ref{tab:table_na}), from $f = 1.20$ for the strongest line ($\lambda 1260.42$) which is clearly saturated, to $f = 0.00245$ for $\lambda 1808.01$; accordingly, it is the weakest, apparently unsaturated (see bottom right panel of Figure \ref{fig:ISM_lines_1}), line that gives the highest measure of $N$. There is mild saturation in the cores of the Si\,{\sc iv} doublet lines (right panel of Figure \ref{fig:fig_nav_1}), although out in the wings there is better agreement between the values of $N_{\rm a}(v)$ indicated by each line. Again, it is the weaker line which probably returns the more reliable estimate of the column density of Si\,{\sc iv}. Thus, focusing on the first ions in Table~\ref{tab:table_na}, we consider the values of $N$ deduced for Si\,{\sc ii}, S\,{\sc ii} and Ni\,{\sc ii} to be the most trustworthy, given that for each of these species our data include apparently unsaturated transitions. From the AOD method, Al\,{\sc iii} $\lambda$1854 also shows mild saturation.  There may well be some mild saturation affecting the single Al\,{\sc ii} and Fe\,{\sc ii} absorption lines in the present set; without additional transitions we cannot estimate the severity of the upward corrections (if any) to be applied to the values of $N$(Al\,{\sc ii}) and $N$(Fe\,{\sc ii}) listed in the Table. The tabulated value of $N$(C\,{\sc ii}) is an uninformative lower limit because $\lambda 1334.53$ is strongly saturated.

Inspection of Table~\ref{tab:table_na} also shows that our errors $\delta N$ (and presumably $\delta$EW) are evidently underestimated. This can most readily be appreciated by realizing that the six independent determinations of $N$(Ni\,{\sc ii}) differ from one another by much more than may be expected on the basis of the quoted errors. In this case, hidden saturation does not appear to be the culprit, since there is no trend of increasing $N$ with decreasing $f$-value. Our Monte Carlo approach to determining the errors, which included random realizations of the continuum level, seems robust. The most plausible explanation, in our view, is that these weak lines are blended to different degrees with photospheric (i.e. stellar) absorption features which are difficult to resolve from interstellar absorption and thus fully account for (as the work of \citealt{Rix+2004} makes clear). As this problem may affect to some extent all of the interstellar features considered, we take the pragmatic approach of adopting the dispersion between the six $N$(Ni\,{\sc ii}) values, $\pm 50$\%, as a more conservative estimate of $\delta N$ for all of the ion column densities.

\subsection{Neutral Hydrogen Column Density from Ly$\alpha$}
\label{sec:Lya}

In the upper panel of Figure \ref{fig:lya} we reproduce the portion of the ESI spectrum encompassing the Ly$\alpha$ line. As can be seen, this region is a complex
blend of several spectral features. In addition to Ly$\alpha$, there are a number of 
metal absorption lines (stellar and interstellar) at wavelengths close to that 
of Ly$\alpha$ (Si\,{\sc ii}~$\lambda\lambda 1190, 1193$; N~\,{\sc i}~$\lambda\lambda\lambda 1199.5, 1200.2, 1200.7$; Si\,{\sc iii}~$\lambda 1206$; N\,{\sc v}~$\lambda\lambda 1238, 1242$). These lines span velocities $\Delta v \geq 1500$\,km~s$^{-1}$ (see Figure \ref{fig:ISM_lines_avg}); together with the ever-present Ly$\alpha$ forest, they make the interpretation of the Ly$\alpha$ line itself somewhat problematic.

In this task, we are aided by the availability of the KCWI integral field spectroscopy of James et al.\ (in preparation), which allows us to explore how this spectral region changes as one moves along the arc of the lensed image. Inspection of the KCWI data cube reveals that the Ly$\alpha$ line itself is a blend of absorption and emission.
There are regions of the arc where the Ly$\alpha$ emission is at a minimum, and others where we only see emission. An example of the former is reproduced in the 
lower panel of Figure \ref{fig:lya}; at this location, we see a clear damped profile indicative of a high column density of neutral hydrogen
$N$(H\,\textsc{i})$ = 3.5 \times 10^{21}$\,cm$^{-2}$
([$\log N$(H\,{\sc i})/cm$^{-2}] = 21.54$). 
A damped profile with a somewhat lower column density
[$\log N$(H\,{\sc i})/cm$^{-2}] = 21.4 \pm 0.1$ provides a reasonable fit 
to the absorption+emission blend of Ly$\alpha$ in
the averaged spectrum captured by the ESI slit. We adopt this value
as our best estimate of the neutral hydrogen column density. 
Note, in this respect, that the 
metal absorption lines in the KCWI spectrum where Ly$\alpha$
is in absorption are consistent within the noise with those
in the ESI spectrum (compare the two panels of Figure \ref{fig:lya}),
lending support to our procedure for establishing the neutral hydrogen column density.
It is interesting that the value of $N$(H\,\textsc{i}) in J1059 is one of the highest so far encountered in star-forming galaxies at $z = 2$--3, and is at the upper end of the distribution of values measured in damped Ly$\alpha$ systems \citep[DLAs;][]{Noterdaeme2014}. 

{The column density derived for J1059 is comparable to the values routinely encountered in Gamma-Ray Burst (GRB) DLAs (e.g. Jakobsson+2006,  Krühler+2013, Bolmer+2019). GRBs occur in star-forming regions within their host galaxies and probe the hydrogen density only along the GRB line of sight. It is remarkable that similarly high values of $N$(H\,\textsc{i}) apply to J1059, even though 
it is an extended object and the ESI slit averages the absorption along thousands of sightlines to OB stars within the $\sim 2$--3\,kpc physical scale of the source
(Figure~\ref{fig:J1059_LensModel}). James et al. (in prep) will further investigate the differential hydrogen column density in the galaxy.}

\section{Chemical Composition of the Interstellar Gas}
\label{sec:chemical_comp}

In principle, having determined the column densities of five elements, from Al to Ni, as well as that of H, we are now in a position to attempt to measure the chemical composition of the interstellar gas in J1059. However, before proceeding we must sound several notes of caution. First, the background continuum against which we see absorption is not a point source but is a composite of many sightlines to the spatially extended starburst (right panel of Figure \ref{fig:J1059_LensModel}); furthermore, we know from the integral field observations of James et al.\ (in preparation) that there are variations in the absorption/emission mix along the ESI slit. As pointed out earlier, what we measure are \textit{average} quantities for the regions encompassed by the spectrograph slit, but the mean residual intensity in an absorption line wavelength bin is only the same as the mean optical depth in that bin in the optically thin regime. This could potentially lead us to underestimate the ion column densities, particularly for species where only one absorption line is available to us, so that hidden line saturation cannot be assessed with the AOD method (e.g.\ Al\,{\sc ii}~$\lambda 1670$ and Fe\,{\sc ii}~$\lambda 1608$).

Second, there are the usual concerns with neglecting potential ionization corrections and dust depletions, which we do not have sufficient data here to estimate quantitatively. The very high neutral hydrogen column density would suggest that ionization corrections might be small \citep{Vladilo2001} but, on the other hand, we do know that there is ionized gas (traced by Al\,{\sc iii}, C\,{\sc iv}, Si\,{\sc iv} and N\,{\sc v}) at the same velocities as the first ions that are the major ionization stages of their elements in H\,{\sc i} regions (see Figure \ref{fig:ISM_lines_avg}). As for depletions, the finding that the continuum of J1059 is reddened by dust assures that, for refractory elements at least, such corrections may well be important.

Finally, for the reasons explained in Section~\ref{sec:MetalLines}, the derivation of realistic uncertainties in the measures of ion column densities is not straightforward. Despite all of these caveats, it is still worthwhile examining the conclusions that can be drawn from the data in Table~\ref{tab:table_na}.

In Table~\ref{tab:table_xh} we collect element abundances in the interstellar gas of J1059 and compare them to the solar composition from \cite{Asplund_Grevesse2009}. The abundances of the two $\alpha$-capture elements covered by our data, Si and S, are similar at $\sim 1/4$ solar within the errors: [Si, S/H]$_{\rm J1059} \sim -0.65$. If we have been pessimistic in assigning errors of $\pm 50$\% to the ion column densities, the true abundance in the gas may be closer to the S value, $\sim 1/3$ solar, with Si showing some mild depletion onto dust \citep[among the elements considered here S has the least affinity for dust---see][]{Jenkins2009, JenkinsWallerstein2017}. Alternatively, as pointed out by \cite{Jenkins2009}, it may be the case that some of the singly ionized S is located in H\,{\sc ii} gas, given that only photons with energies greater than 23.4\,eV (significantly higher than the 13.6\,eV required to ionize H) can produce S$^{++}$. 

The two Fe-peak elements considered here, Fe and Ni, are both less abundant than Si and S. This could be understood as a combination of dust depletion (both are refractory elements easily incorporated into dust grains) and an intrinsically lower than solar Fe-peak to $\alpha$-capture element ratio which is now acknowledged to be a common feature of galaxies that are rapidly forming stars \citep[e.g.][]{Steidel+2016, Strom+2018, Sanders+2020}.

More difficult to interpret are the \textit{relative} abundances of Fe and Ni, with the former apparently four or five times less abundant than the latter. Such a difference is unexpected: these two elements usually track each other closely in stars \citep[e.g.][]{Reddy+2003}, and are depleted to similar extent in the interstellar medium of the Milky Way and the Small Magellanic Cloud \citep{Jenkins2009, JenkinsWallerstein2017}.  Possible explanations for this apparent anomaly are that: (a) the column density of Fe\,{\sc ii} has been underestimated if there is significant saturation in the $\lambda 1608$ absorption line, and/or (b) the column density of Ni\,{\sc ii} has been overestimated if there is stellar photospheric contamination of the six Ni\,{\sc ii} absorption lines (Table~\ref{tab:table_na}) which we consider to be exclusively interstellar. (Differential ionization effects are unlikely to account for the factor of $\sim 5$ difference given that the ionization potentials of Fe$^+$ and Ni$^+$ differ by less than 2\,eV.) 

\begin{table}
	
	\caption{Chemical abundances}
	\label{tab:table_xh}
	\begin{center}
	\addtolength{\leftskip}{-0.8cm}
	\begin{threeparttable}
	 
	%\makebox[\linewidth]{\scriptsize%
		\begin{tabular}{lcccc} % four columns, alignment for each
		\hline
		Ion & $\log\,(N/\rm cm^{-2})$  &  $\rm log(X/H)$ & $\rm log(X/H)_{\odot}^{\tnote{a}}$ & $\rm [X/H]_{J1059}^{\tnote{b}}$ \\
		\hline
		H\,{\sc i}   & $21.40 \pm 0.10$   &  ...& ... &... \\
        \\
        Al\,{\sc ii} & $13.97^{+0.18}_{-0.30}$ & $-7.43^{+0.19}_{-0.35}$ & $-5.55$ & $-1.88^{+0.19}_{-0.35}$\\
        \\
        Si\,{\sc ii} & $16.06^{+0.18}_{-0.30}$ & $-5.34^{+0.19}_{-0.35}$ & $-4.49$ & $-0.85^{+0.19}_{-0.35}$\\
		\\
        S\,{\sc ii}  & $16.02^{+0.18}_{-0.30}$ & $-5.38^{+0.19}_{-0.35}$ & $-4.88$ & $-0.50^{+0.19}_{-0.35}$\\
        \\
		Fe\,{\sc ii} & $15.26^{+0.18}_{-0.30}$ & $-6.14^{+0.19}_{-0.35}$ & $-4.50$ & $-1.64^{+0.19}_{-0.35}$\\
		\\
		Ni\,{\sc ii} & $14.65^{+0.18}_{-0.30}$ & $-6.75^{+0.19}_{-0.35}$& $-5.78$ &$-0.97^{+0.19}_{-0.35}$\\
		\\
		\hline
	\end{tabular}
		\begin{tablenotes}
	
  \item[a] Solar abundance scale from \citet{Asplund_Grevesse2009}  
  \item[b] $\rm [X/H]_{J1059}=log(X/H)_{J1059}-log(X/H)_{\odot}$
  \end{tablenotes}
\end{threeparttable}
\end{center}
\end{table}

\begin{figure}
	\includegraphics[width=0.5\textwidth]{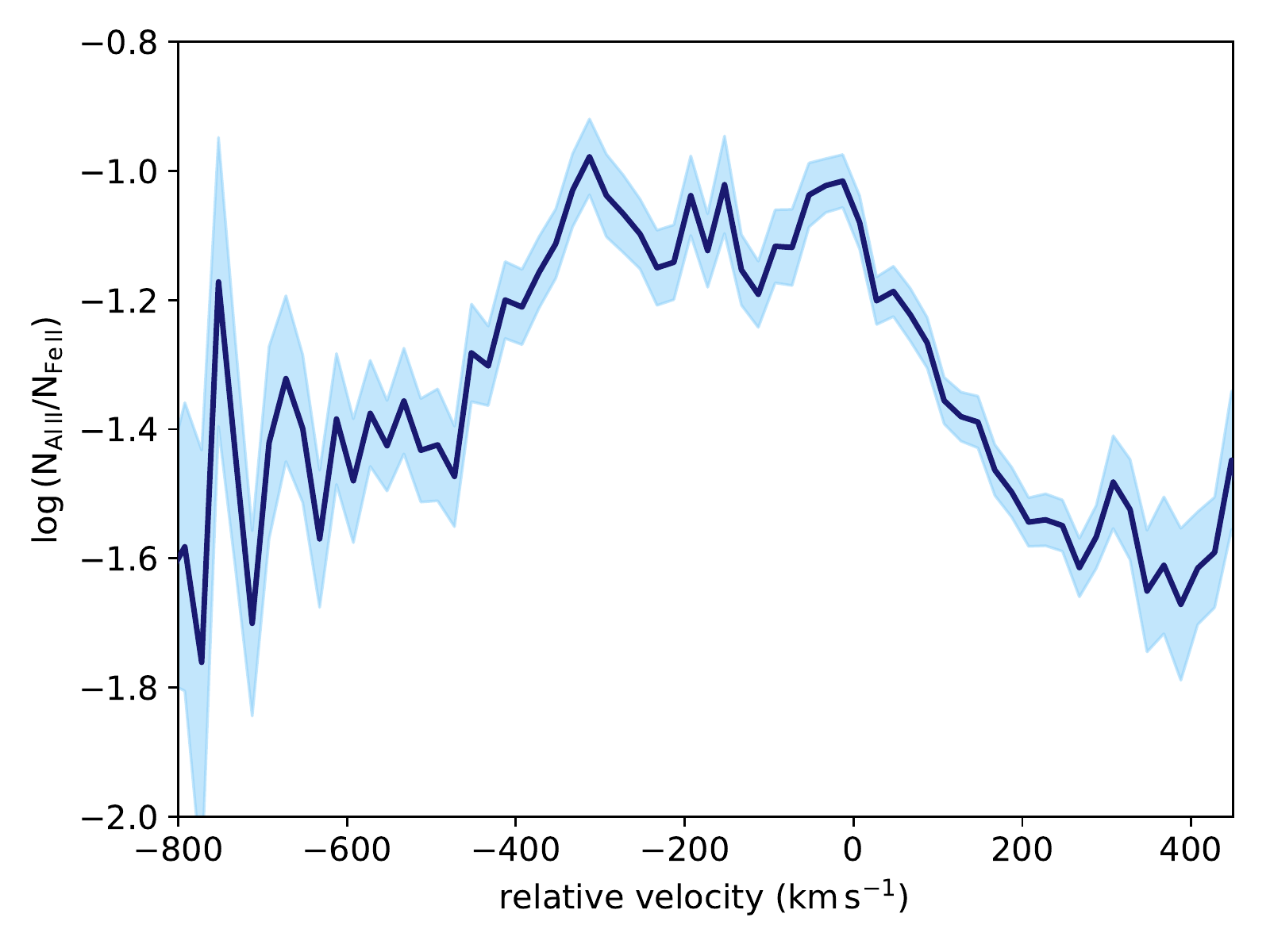}
      \caption{Pixel-by-pixel ratio (and relative error) between $N$(Al\,{\sc ii}) and $N$(Fe\,{\sc ii}) as a function of relative velocity.}
   \label{fig:fig_abd_3}
\end{figure}

We address the first of these options by considering Fe\,{\sc ii}~$\lambda 1144.9379$, the only other Fe\,{\sc ii} line included in our spectrum. This line is recorded at lower S/N than $\lambda 1608$, being close to the short wavelength limit of ESI, and its measurement is complicated by blending with the Ly$\alpha$ forest (which is why it is not included in Table~\ref{tab:table_na}). Nevertheless, comparing its apparent optical depth to that of the weaker $\lambda 1608$ line does indicate some degree of saturation, which is likely to affect both transitions. On the other hand, inspection of the {\sc starburst}99 best fitting spectrum reproduced in Figure \ref{fig:J1059_fit} (see Section~\ref{sec:stellar_pop}) does not support the second option as the main reason for the anomalously low Fe/Ni ratio. 

If Fe\,{\sc ii}~$\lambda 1608$ suffers significant, unresolved, saturation the same is likely to apply to Al\,{\sc ii}~$\lambda 1670$ which has similar apparent optical depth (see Figures~\ref{fig:ISM_lines_1} and \ref{fig:ISM_lines_2}) and this, together with dust depletion, may also explain the low abundance of Al in the gas. However, we note that there appear to be real differences with velocity in the ratio of the apparent optical depths of these two lines, as can be appreciated from Figure \ref{fig:fig_abd_3}. Changes in the ionization balance \citep{Vladilo2001}, degree of depletion, and intrinsic abundance could all be playing a part.

\begin{figure}
	\includegraphics[width=0.5\textwidth]{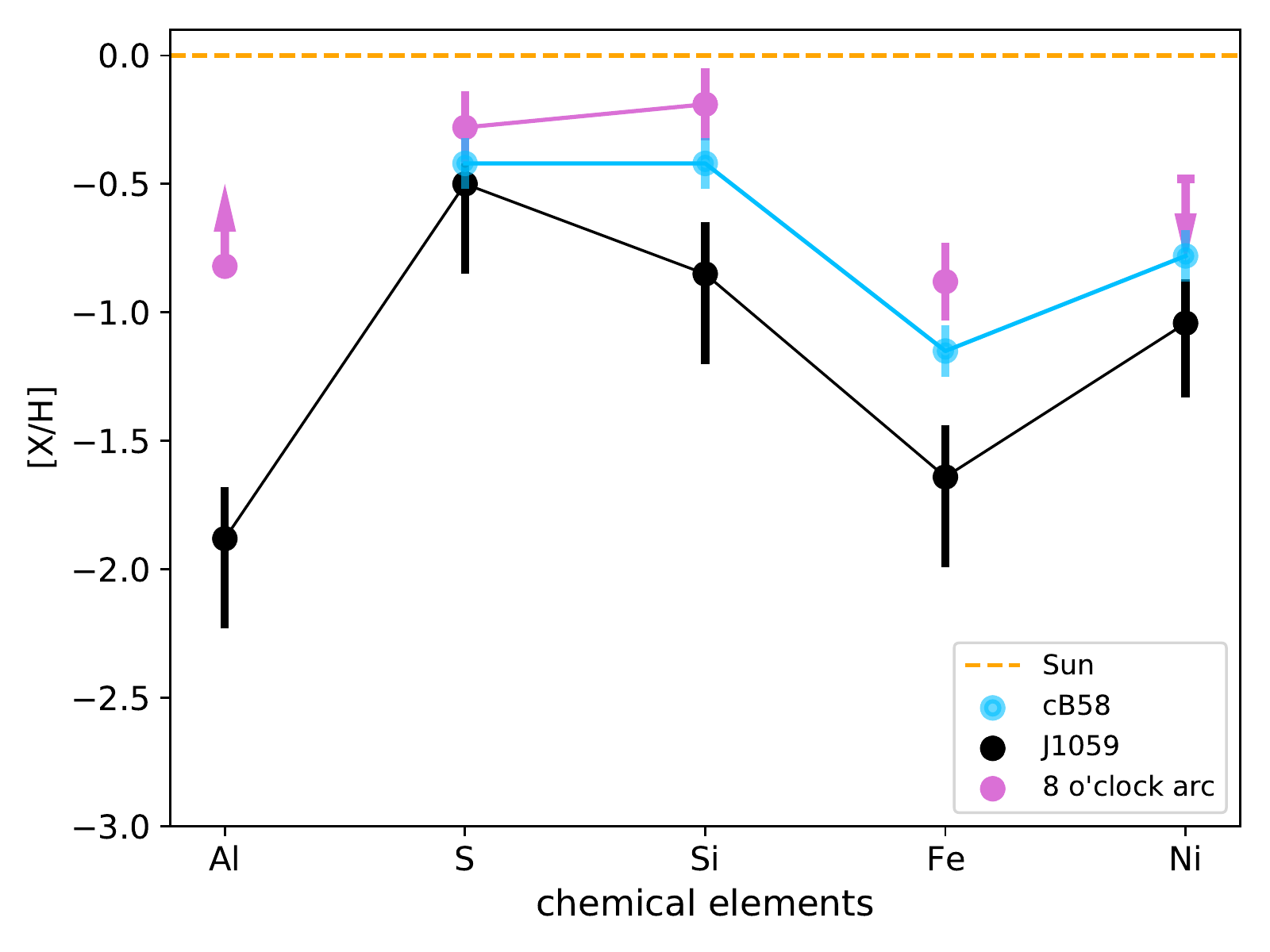}
	   \caption{Element abundances in the interstellar gas of J1059 compared with those in two other well-studied lensed galaxies: MS15-cB58 (cyan) and the 8 o'clock arc (magenta).}
   \label{fig:fig_abd_1}
\end{figure}

In Figure \ref{fig:fig_abd_1}, we compare the pattern of abundances for the five elements considered here with analogous measurements in two other well-studied gravitationally lensed galaxies whose absorption-dominated spectra have allowed the composition of the interstellar gas to be determined: MS1512-cB58 \citep{Pettini+2002} and the 8 o'clock Arc \citep{Dessauges-Zavadsky+2010}. All five elements are less abundant in J1059 than in the other two cases, but it is interesting to note that the difference is least pronounced for S, the element least likely to be incorporated into dust grains. Among the three galaxies, J1059 is also the one showing the highest value of neutral hydrogen column density ($\log N$(H\,{\sc i})/cm$^{-2} = 21.4$), compared to 20.85 in cB58 and 20.57 in the 8 o'clock Arc, further pointing to a combination of line saturation and dust depletion as plausible causes for the much lower abundances of Fe and Al in J1059.

It is also of interest to compare the chemical enrichment in the interstellar gas with our earlier findings regarding the metallicity of the young stellar population, In Section~\ref{sec:stellar_pop}, we concluded that a metallicity $Z_* \simeq 0.004$ is favoured by the comparison of {\sc starburst}99 models with the UV stellar spectrum of J1059, with a $1 \sigma$ range $Z_* = 0.003$--0.0085. The comparison with the data in Table~\ref{tab:table_xh} is made difficult by the uncertainties in both stellar and interstellar abundances, which are not insignificant, and by the fact that the {\sc starburst}99 models are built with the solar \textit{relative} abundance scale which probably does not apply to actively star-forming galaxies, as mentioned earlier. Thus, the "metallicity" of {\sc starburst}99 model spectra is some average of the abundances of many elements, including both alpha (mainly O) and iron-peak (mainly Fe) elements, as well as C and N which we do not measure in the interstellar gas. 
With all these reservations in mind, we note that $Z_* \sim 0.004$ corresponds to $\sim 1/3\, Z_\odot$ \citep{Asplund_Grevesse2009} or $-0.5$\,dex, which agrees with the abundance of S (and Si, particularly if the latter suffers mild depletion) in Table~\ref{tab:table_xh}.\\
\citet{James&Aloisi2018} find that the abundances of S and O are correlated in local galaxies, and if we assume the same correlation we find $\rm 12+log(O/H) = 8.17^{+0.2}_{-0.3}$. 
With this oxygen abundance and our measured hydrogen column density $\log\,N$(H {\sc i})$=21.4\pm0.1$, J1059 falls on the local relationship between $\rm O/H$ and  $N$(H {\sc i}) \citep{James&Aloisi2018}.
{However, local galaxies that have similar oxygen abundance and hydrogen column density are less massive than J1059 by an order of magnitude.} This difference is broadly consistent with the offset of the mass-metallicity relation to lower metallicities at a given stellar mass at higher redshifts (\citealp{Erb+2006, Henry+2013, Steidel+2014, Sanders+2020}), and with the finding that local galaxies matching the excitation properties of $z\sim2$ galaxies are $\sim10$ times less massive than their higher redshift counterparts \citep{Strom+2017}.

%\citet{James&Aloisi2018} show the presence of a clear trend between the gas metallicity and the hydrogen column density for nearby galaxies at $z<0.1$, which suggests galaxies of any mass tend to be characterized by inflows of pristine gas, but only lower mass galaxies with shallower potential wells experience metal loss through gas outflows. From the abundance of S, we derive $\rm 12 + log(O_{S}/H)=8.17^{+0.2}_{-0.3}$ ($\sim 1/3 Z_{odot}$) for J1059, which therefore falls on the trend, matching the nearby low mass ($<10^9M_{\odot}$), high-redshift analogs, NGC 5253 and NGC 4449 (See \citealp{James&Aloisi2018} and \citealp{James+2014} for further details). This suggests that even if on different mass scales, galaxies experiences similar star formation processes at different cosmic epochs.}

\section{Discussion and Conclusions}
\label{sec:conc}

In this paper we have presented ESI observations of the rest-frame UV spectrum of SDSS J1059+4251 (J1059), a star-forming galaxy at $ z=2.8$, gravitationally lensed by a complex of galaxies at $z\sim0.7$. J1059 is very bright ($\rm F814W=18.8\,mag$), due both to its high intrinsic luminosity and the gravitational lensing that provides one of the highest magnification factors observed so far in a galaxy-scale lens ($\mu=31 \pm 3$). The ESI observations therefore provide high S/N ($\sim 30-40$ over the analyzed wavelength range) at high spectral resolution ($\rm FWHM=\,60\, km\,s^{-1}$). 

The aim of this work was to characterize the average properties of J1059, while in a following paper (James et al.\ in prep.) we will focus on their variation on sub-kpc scales.
The relevance of the present project is that the magnification of J1059, coupled with the high resolution of the ESI spectrum, enables us to separate the interstellar features from the stellar ones and to derive properties that would be inaccesible without the aid of the lensing.
Our main findings can be summarized as follows.

\begin{itemize}

     \item {From SED fitting of the \textit{HST} and WISE photometry, we derive a stellar mass $M_*= (3.22\pm 0.20)\,\times10^{10}$ $\rm M_{\odot}$, age $644^{+76}_{-90}$ Myr, $E(B-V)= 0.05\pm 0.012$, and $\rm SFR = 50 \pm 7\, M_{\odot}$~yr$^{-1}$. From the UV spectrum calibrated with the \textit{HST} photometry, we find a UV slope $\beta=-1.61\pm 0.08$, while the rest-frame UV luminosity yields extinction-corrected SFRs of 70--90\,$M_{\odot}$~yr$^{-1}$ depending on the calibration used (see Section \ref{sec:general_prop});}
     
     \item We fit the whole ESI UV spectrum, including the stellar continuum and stellar features with {\sc starburst}99 and BPASS models, deriving a subsolar stellar metallicity of $Z\sim0.15-0.5\,Z_{\odot}$ {(Section \ref{sec:stellar_pop})}; 
    
    \item We study the ISM line profiles, finding evidence of large scale outflows powered by the starburst and extending up to $\rm \sim - 1000\,km\,s^{-1}$. We also find an absorption component at positive velocities which suggests the presence of inflowing gas {(Section \ref{sec:kinematics})};
    
     \item We analyze the pattern of chemical abundances in J1059 deduced from fitting individual ISM absorption features. We find that Fe-peak elements (Fe and Ni) are less abundant than $\alpha$-capture elements (Si and S). However, these trends can be affected by dust depletion, which seems to be higher in J1059 compared to other lensed galaxies studied so far. 
     We find that the best-fit stellar metallicity $Z_* \sim 0.004$ is in agreement with the abundance of S (and Si, particularly if the latter suffers mild depletion) in the gas {(Section \ref{sec:MetalLines}).}
     
\end{itemize}

J1059 is a typical example of a star-forming galaxy at $z\sim3$, with a stellar mass and a star formation rate matching the star-forming main sequences of galaxies at similar redshifts (e.g. \citet{Santini+2017}). This consistency suggests that J1059 is not experiencing a short-lived starburst or merger event, and fits in the picture that merger-enhanced SFRs are relatively unimportant in $z\sim2$ galaxies, as shown by \citet{Rodighiero+2011}.

The stellar mass and subsolar stellar metallicity we derived for J1059 fall on the stellar mass-stellar metallicity relation found by \citet{Cullen+2019} for VANDELS \citep{McLure+2018} galaxies with $M_*>10^{10}\,M_{\odot}$ in the redshift range $2<z<5$. Similarly to our approach, these authors fit the galaxy rest-frame UV spectra with {\sc starburst}99 models. Our stellar mass and stellar metallicity values also fall onto the stellar mass-stellar metallicity relation derived by \citet{Calabro+2020} by means of the photospheric absorption indices defined by \citet{Rix+2004}. Photospheric indices have the advantage of being relatively immune to overall changes in the continuum shape and therefore are less affected by the degeneracy between age and dust. Our metallicity result from the fit with {\sc starburst}99 and BPASS models is also consistent with that obtained by \citet{Sommariva+2012} for an ultraviolet-selected AMAZE \citep{Mannucci+2009} galaxy at $z\sim3.4$ with mass comparable to ours. In this context, J1059 represents additional evidence that subsolar metallicities are common among galaxies at $2<z<5$. Our findings also show that different methods (i.e.\ full-spectrum fitting and photospheric absorption indices) are consistent in predicting the stellar properties of high redshift galaxies. It is also worth noting that, even though our full-spectrum analysis can be affected by the age-dust degeneracy, its agreement with the results from the photospheric indices points towards the robustness of our results. 

Since bright star-forming disk galaxies contain most of the H {\sc i} mass in the nearby Universe, it has always seemed likely that these galaxies and their high-redshift progenitors would be the origin of the Damped Ly$\alpha$ systems (DLAs) seen in background quasar spectra. The current view, based on both theoretical \citep{Pontzen+2008, Berry+2016, DiGioia+2020} and observational 
\citep{Krogager+2017, Krogager+2020} results is that DLAs are in fact a broad class of galaxies selected by H\,{\sc i} cross-section spanning a range of 2--3 orders of magnitude in both mass and metallicity. Star-forming galaxies with absorption-dominated spectra tend to lie at the high value ends of the distributions of $N$(H {\sc i}) and $Z$ spanned by the general DLA population, and in this respect the data reported here for J1059 fit this trend.

The stellar properties of J1059 we have discussed so far confirm some of the evolutionary trends that have been observed in high redshift galaxies in the past decades. This also holds for dust extinction, which we parameterized through the UV slope $\beta$. The value $\beta=-1.61 \pm0.08$ that we derive is consistent, within the uncertainties, with the median value $\langle\beta\rangle=-1.70\pm0.55$ found by \citet{Pilo+2019} in a sample of 517 $z\sim3$ bright ($-24<M_{1600}<-21$) COSMOS \citep{Taniguchi+2007} star-forming galaxies. Moreover, our values of redshift and $\beta$ fall on the best fit $z-\beta$ relation found by \citet{Calabro+2020}. They explored the $\beta$ slope of $> 500$ star-forming galaxies at redshifts $2<z<5$ extracted from the VANDELS \citep{McLure+2018} survey and found that $\beta$ increases on average from $-1.98$ at $z\sim4.1$ to $-1.59$ at $z\sim2.6$. Our result therefore matches the strong evolution in $\beta$ seen at these cosmic epochs \citep{Pannella+2015}.

Galaxy formation involves a continuous competition between gas cooling and accretion on the one hand, and feedback-driven heating and/or mass outflows on the other. Outflows, which are locally detected only in starburst galaxies (e.g. \citealp{Heckman+2000}, \citealp{Martin2005}, \citealp{Heckman+2015}) are very common at higher redshifts ($z>0.5$) among the general star-forming galaxy population. However, inflows of accreting cold gas at high redshift are very elusive and difficult to observe, since, as suggested by \citet{Steidel+2010}, they are often obscured by outflows or by absorption from the galaxy’s ISM. For example, \citet{Martin2005} conducted an ESI study of the interstellar Na\,{\sc i} D lines in 18 local ultraluminous infrared galaxies, finding evidence of outflows in 15 cases and of inflow in only one case. Also, theoretical studies have failed to fully predict the observational properties of inflows. The predictions are in fact rather model-dependent for both absorption lines and Ly$\alpha$ emission. Moreover, the simulations which predict cold accretion generally do not account for interstellar gas that may have been carried to large galactocentric radii by outflows, nor for the scattering of Ly$\alpha$ photons before escaping the galaxy (see \citealp{Steidel+2010}). The investigation of observed velocity profiles is therefore considered one of the most effective ways to capture signatures of outflowing and cold inflowing gas filaments, which appear as blueshifted and redshifted components in the ISM absorption lines. These results are particularly clear in the case of J1059 thanks to the magnification and the high spectral resolution of the ESI spectrum, and offer still rare evidence of the presence of inflows in high redshift galaxies. In particular, we may be viewing J1059 along a line of sight that captures a cold filament of dense  material accreting onto the galaxy.

Thanks to the high data quality, we are able to derive chemical abundances from individual absorption lines. This is remarkable considering that (to our knowledge) there are very few (lensed) galaxies with measured abundance patterns at these redshifts (cB58, \citealp{Pettini+2002}; the 8 o'clock arc, \citealp{Dessauges-Zavadsky+2010}). In particular, the picture emerging from our work is consistent with rapid star formation, where the $\alpha$ elements produced by the short-lived Type II Supernovae are more abundant than the Fe-peak elements produced by later Supernovae Ia events. It is important to point out that the high hydrogen density of J1059 suggests a higher level of saturation and dust depletion than other lensed galaxies at similar redshifts, and these two factors may be playing a role in shaping the observed trends. However, this same trend of overabundant $\alpha$-capture elements has been observed in other lensed galaxies (\citealt{Pettini+2002}, \citealt{Dessauges-Zavadsky+2010}),
and indeed appears to be common to most star-forming galaxies at $z = 2$--3 (\citealt{Steidel+2016}, \citealt{Strom+2018}).

From these results, we can conclude that a full understanding of the interplay of different factors on the observed chemical abundances still needs to be achieved and that our ability to perform exhaustive studies on the star formation histories of galaxies at very early epochs is still limited. 
However, spatially resolved studies of lensed galaxies afforded by  high-sensitivity IFU instruments such as MUSE \citep{Bacon+2010}, KCWI \citep{Morrissey+2018}, and the upcoming NIRSpec \citep{Bagnasco+2007} on the \textit{James Webb Space Telescope} will be useful to overcome complications due to the blending of sightlines.
Larger samples of gravitationally lensed galaxies, especially the most highly magnified ones, are also needed.
In this context, future surveys such as the Vera Rubin Observatory Legacy Survey of Space and Time are expected to discover $\gtrsim100,000$ new lenses \citep{Collett+2015},  constituting a vast database for high-redshift studies such as the one we have presented here.

\begin{acknowledgements}

The authors thank the anonymous referee for very helpful comments which have improved the manuscript. The authors thank Naveen Reddy for assisting with the SED fitting and  Ed Jenkins for helpful advice on the Ni\,{\sc ii} $f$-values. This work was supported by a NASA Keck PI Data Award, administered by the NASA Exoplanet Science Institute. A.C. and D.K.E. are supported by the US National Science Foundation (NSF) through the Faculty Early Career Development (CAREER) Program grant AST-1255591 and the Astronomy \& Astrophysics grant AST-1909198. B.L.J thanks support from the European Space Agency (ESA). Data presented herein were obtained at the W. M. Keck Observatory from telescope time allocated to the National Aeronautics and Space Administration through the agency's scientific partnership with the California Institute of Technology and the University of California. The Observatory was made possible by the generous financial support of the W. M. Keck Foundation. The authors wish to recognize and acknowledge the very significant cultural role and reverence that the summit of Mauna Kea has always had within the indigenous Hawaiian community. We are most fortunate to have the opportunity to conduct observations from this mountain.
\end{acknowledgements}

\facilities{Keck II (ESI), HST (WFPC3)}
\software{Astropy \citep{Astropy2013,Astropy2018},  PySynphot \citep{pysynphot}, FSPS \citep{Conroy+2009,Conroy+2010}, python-FSPS \citep{python-fsps}, Prospector \citep{Johnson+2020}, emcee \citep{emcee}}

\bibliographystyle{aasjournal}
\bibliography{bibliography.bib}

%%%%

%%%%

\end{document}